\documentclass[aps,pra,showpacs,superscriptaddress,tightenlines,twocolumn,lengthcheck]{revtex4-2}
\usepackage{amssymb}
\usepackage{amsmath}
\usepackage{graphicx}
\usepackage{txfonts}
\usepackage{color}

\usepackage{amsfonts}
\usepackage[colorlinks,urlcolor=blue,linkcolor=blue,citecolor=blue]{hyperref}

\begin{document}

\title{Deterministic generation of multi-photon bundles in a quantum Rabi model}
\author{Cheng Liu}
\affiliation{Key Laboratory of Low-Dimensional Quantum Structures and
Quantum Control of Ministry of Education, Key Laboratory for Matter
Microstructure and Function of Hunan Province, Department of Physics and
Synergetic Innovation Center for Quantum Effects and Applications, Hunan
Normal University, Changsha 410081, China} 

\author{Jin-Feng Huang}
\email{Corresponding author: jfhuang@hunnu.edu.cn}
\affiliation{Key Laboratory of Low-Dimensional Quantum Structures and
Quantum Control of Ministry of Education, Key Laboratory for Matter
Microstructure and Function of Hunan Province, Department of Physics and
Synergetic Innovation Center for Quantum Effects and Applications, Hunan
Normal University, Changsha 410081, China} 

\author{Lin Tian}
\affiliation{School of Natural Sciences, University of California, Merced, California 95343, USA} 

\begin{abstract}
Multi-photon bundle states are crucial for a broad range of applications such as quantum metrology, quantum lithography, quantum \textcolor{black}{communication}, and quantum biology. Here we propose a scheme that generates multi-photon bundles via virtual excitations in a quantum Rabi model. 
Our approach utilizes a $\Xi$-type three-level atom, where the upper two levels
are coupled to a cavity field to form a quantum Rabi model with ultrastrong coupling, and the transition between the lower two levels is driven by two sequences of Gaussian pulses. We show that the driving pulses induce \textcolor{black}{deterministic} emission of \textcolor{black}{multiple} photons from \textcolor{black}{the eigenstates} of the quantum Rabi model via the stimulated Raman adiabatic passage technique, and hence can create bundles of \textcolor{black}{multiple}  photons \textcolor{black}{on-demand} in the cavity output field. 
We \textcolor{black}{calculate} the generalized second-order correlation functions of the output photons, which reveal that the emitted photons form antibunched multi-photon \textcolor{black}{bundles}.
\end{abstract}
\date{\today}
\maketitle

\section{Introduction\label{introduce}}
Light-atom interaction plays a key role in quantum optics and quantum information. 
The quantum Rabi model (QRM), which describes the interaction of a bosonic mode with a two-level atom, \textcolor{black}{has been widely studied} in quantum optics and quantum science. 
With the progress in experimental technology in the last few decades, the \textcolor{black}{strength of the} light-atom interaction can now exceed 10\% of the light
frequency \textcolor{black}{and} the atomic transition frequency~\cite%
{Niemczyk2010,Forn2010,Schwartz2011,Benz2016,Askenazi2017,Barachati2018}, which is denoted as the ultrastrong coupling regime, or can even be comparable to the light or atomic frequency, which is called the deep-strong coupling regime~\cite{Yoshihara2017,Bayer2017,Yoshihara2018}. 
The ultrastrong coupling regime can be implemented with various systems including the superconducting quantum devices~\cite%
{Niemczyk2010,Forn2010,Yoshihara2017,Yoshihara2018}, intersubband polaritons~%
\cite{Askenazi2017}, Landau polaritons~\cite%
{Bayer2017,Muravev2011}, organic molecules~\cite{Schwartz2011,Barachati2018}%
, and optomechanical systems~\cite{Benz2016}. Many interesting phenomena resulted from the ultrastrong or deep-strong coupling have been studied or demonstrated in these systems, such as vacuum degeneracy~\cite{Nataf2010}, photon blockade~\cite{Ridolfo2012}, few-photon scattering~\cite%
{Wang2012,Shi2018}, quantum phase transition~\cite{Hwang2015,Chen2021,Cai2021,Chenye2022}, multi-photon Rabi oscillation~\cite{Garziano2015,Ma2020}, \textcolor{black}{controllable} counter-rotating interaction~\cite{Huang2015,Huang2017,Huang2020, Huang2022}, and few-photon emission~\cite%
{Stassi2013,Huang2014,Cirio2016}.
\textcolor{black}{With} ultrastong coupling, the rotating-wave approximation (RWA) fails, and the ground state of the QRM carries virtual photons that cannot be \textcolor{black}{directly emitted or} detected. Some schemes
such as spontaneous emission~\cite{Stassi2013}, stimulated emission~\cite%
{Huang2014}, and electroluminescence~\cite{Cirio2016} have been proposed to generate real photons by 
converting the virtual photons in the ground state of the QRM.

Recently, multi-quanta physics has attracted enormous interest because of its potential applications \textcolor{black}{ in quantum information sciences}. In particular, the emission of multi-photon bundles~\cite{Munoz2014} has important
applications in the generation of new light source~\cite{Walther2006,Brien2009}, quantum
metrology~\cite{Giovannetti2004,Giovannetti2006}, quantum lithography~\cite{Angelo2001},
quantum communication~\cite{Kimble2008}, quantum biology~\cite%
{Ball2011,Sim2012}, and medical applications~\cite%
{Denk1990,Horton2013}. \textcolor{black}{Multi-photon bundle describes the emission of multiple photons in an antibunched bundle} .
People have investigated the generation of
multi-photon bundles in various setups, such as Rydberg atomic ensembles~%
\cite{Bienias2014,Maghrebi2015}, Kerr cavity systems~\cite%
{Liao2010,Miranowicz2013}, multi-level atomic systems~\cite%
{Dousse2010,Ota2011,Muller2014,Chang2016}, cavity quantum electrodynamics
(QED) systems~\cite{Munoz2014,Strekalov2014,SanchezMunoz2018,Bin2021,Deng2021}, superconducting circuits~\cite{Ma2021}, and
waveguide-QED systems~\cite{Gonz2015,Douglas2016, Gonz2017}. 
However, because the high-order processes of single-photon transition are weak, multi-photon bundle emission is challenging to achieve experimentally. 

Here, we propose an efficient scheme to generate multi-photon bundles via the stimulated Raman adiabatic passage (STIRAP) technique \textcolor{black}{deterministically}~\cite{Bergmann1998,Vitanov2017, Liu2022}, where multiple photons \textcolor{black}{from the eigenstates}  of a QRM are \textcolor{black}{created} as cavity photons, and subsequently released from the cavity in a bundle.  In this scheme, the multi-photon bundle is emitted on demand, controlled by external \textcolor{black}{STIRAP} driving pulses. 
To be specific, we study a $\Xi$-type atom with the upper two levels of the atom coupled to a
cavity field with ultrastrong coupling, while the transition between the lower two levels of the atom is driven by two sequences of external Gaussian pulses. By choosing appropriate resonance conditions,  we can create a $\Lambda$-type three-level system, where one of the lower \textcolor{black}{levels} contains even \textcolor{black}{or odd} number of cavity photons. By applying the STIRAP technique, the system can be deterministically prepared to the lower level with \textcolor{black}{desired number of} cavity photons, which will be emitted as multi-photon bundle via cavity dissipation. Using the quantum trajectory technique, we demonstrate the dynamical emission of
multi-photon bundles from the cavity. We also calculate the standard and generalized second-order correlation functions, which shows the antibunching nature of the emitted multi-photon bundles.  
Our scheme connects virtual photons in the QRM with on-demand, efficient multi-photon bundle emission, and provides a new mechanism for the deterministic generation of multi-photon \textcolor{black}{source}.

\begin{figure}[tbp]
	\center
	\includegraphics[clip, width=8.3cm]{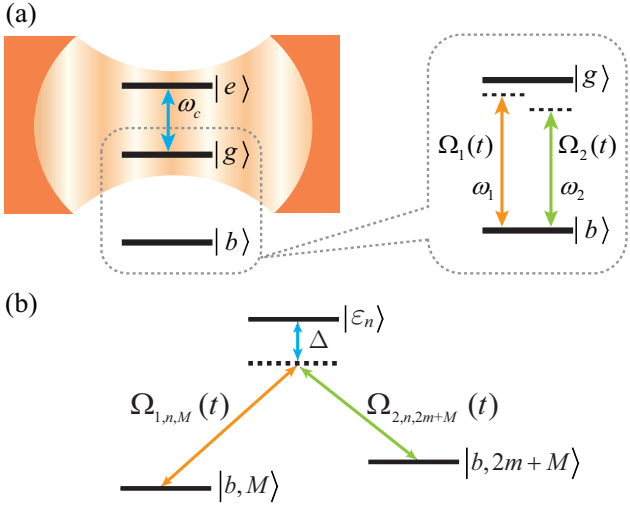}
	\caption{(Color online)~(a) Schematic of the system with a $\Xi$-type
		three-level atom coupled to a cavity mode through the upper two
		levels $\left\vert e\right\rangle $ and $\left\vert g\right\rangle $. The transition between the lower two levels $\left\vert g\right\rangle $ and $\left\vert b\right\rangle 
		$ is driven by two external driving fields with driving frequency $%
		\protect\omega_{l}\left(l=1,2\right)$ and time-dependent
		driving amplitude $\Omega _{l}\left( t\right)$. (b) Energy structure of the effective 
		$\Lambda$-type three-level system. The initial state \textcolor{black}{$\left\vert
			b,M\right\rangle$}  and the final state \textcolor{black}{$\left\vert b,2m\!+\!M\right\rangle$} are
		coupled by the driving fields with the effective coupling strengths \textcolor{black}{$
			\Omega_{1,n,M}\left( t\right)$} and \textcolor{black}{$\Omega_{2,n,2m+M}\left( t\right)$} via the
		\textcolor{black}{eigenstate $\left\vert\protect\varepsilon_{n}\right\rangle$} of the quantum
		Rabi model, \textcolor{black}{where \textcolor{black}{$M\!=\!0$ or 1}}. The detuning $\Delta$ is the difference between the
		driving frequency $\protect\omega_{1}$ ($\protect\omega_{2}$) and the
		transition frequency from \textcolor{black}{$\left\vert b,M\right\rangle$ ($\left\vert
			b,2m\!+\!M\right\rangle$) to $\left\vert \protect\varepsilon_{n}\right\rangle$}.}
	\label{Fig1_model}
\end{figure}

\section{Model}~\label{model_hamiltonian}
We consider a $\Xi $-type three-level atom where the upper two levels $
\left\vert e\right\rangle $ and $\left\vert g\right\rangle $ are
coupled to a cavity mode with ultrastrong coupling, the transition between the lower two levels $\left\vert g\right\rangle $ and $\left\vert b\right\rangle $ is driven by two external driving fields with driving frequency $\omega_{l}\left( l=1,2\right) $ and composed of consecutive Gaussian wave packets. The frequency difference between the lower levels $\left\vert g\right\rangle -\left\vert b\right\rangle $ is much greater
than both the cavity frequency $\omega _{c}$ and the \textcolor{black}{transition frequency} between the upper levels $
\left\vert e\right\rangle -\left\vert g\right\rangle $ so that the bottom level $\left\vert
b\right\rangle $ is not coupled to the cavity mode and the driving fields do not induce transition between the upper levels, as shown in~\ref{Fig1_model}(a).
The system Hamiltonian can be written as ($\hbar =1$) 
\begin{eqnarray}
	H\left( t\right) &=&\sum_{s=e,g,b}\omega _{s}\left\vert s\right\rangle
	\left\langle s\right\vert +\omega _{c}a^{\dag }a+\lambda (a+a^{\dag })\left(
	\left\vert e\right\rangle \left\langle g\right\vert +\left\vert
	g\right\rangle \left\langle e\right\vert \right)  \notag \\
	&&+\sum_{l=1}^{2}\left[ \Omega _{l}\left( t\right) \cos \left( \omega
	_{l}t\right) \right] \left( \left\vert b\right\rangle \left\langle
	g\right\vert +\left\vert g\right\rangle \left\langle b\right\vert \right)
	\label{H}
\end{eqnarray}%
with \textcolor{black}{the} time-dependent driving amplitude 
\begin{equation}
	\Omega _{l}\left( t\right) =\Omega _{l}\sum_{k=0}^{\infty }\exp \left[ -%
	\frac{\left( t-t_{l}-kT_{1}\right) ^{2}}{T^{2}}\right], \hspace{0.5cm}%
	(l=1,2).
\end{equation}%
Here, $a$ $(a^{\dag })$ is the annihilation (creation) operator of the cavity
mode with resonance frequency $\omega _{c}$, $\omega _{s}$ is the
frequency for the energy level $\vert s\rangle$ $\left( s=e,g,b\right) $,  and $\lambda $ is the atom-cavity
\begin{figure}[tbp]
	\center
	\includegraphics[width=8.3cm]{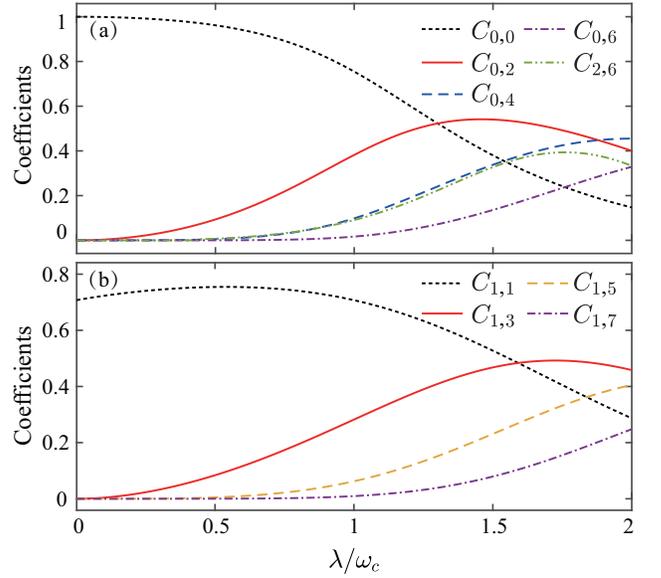}
	\caption{(Color online) \textcolor{black}{Coefficients (a)  of $
			\left\vert g,2m\right\rangle$  in the eigenstate 
			$\left\vert\protect\varepsilon_{n}\right\rangle$ ($n$=0 and 2) and (b)  of $
			\left\vert g,2m\!+\!1\right\rangle$ in the first excited state $\left\vert\protect\varepsilon_{1}\right\rangle$} of $H_{\rm R}$ as a function of the ratio $\protect\lambda/\protect\omega_{c}$ for
		the resonant case $\protect\omega_{c}=\protect\omega_{e}-\protect\omega_{g}$.}
	\label{fig_7_virtula_photon_Am}
\end{figure}
\noindent  coupling strength. The parameters $%
\Omega _{l}$ and $T$ denote the maximum amplitude and the width of the Gaussian pulses, respectively,  $t_{l}$ $(l=1,2)$ is the time of the maximum value of the first Gaussian pulse \textcolor{black}{in each pulse sequence}, $k$ is an integer labelling the pulses in the driving field, and $T_{1}$ is the time interval between consecutive Gaussian pulses.

The Hamiltonian of the upper two levels and the cavity mode 
$H_{\rm R}=\omega _{e}\vert
e\rangle \langle e\vert +\omega _{g}\vert
g\rangle \langle g\vert +\omega _{c}a^{\dag }a\left(
\vert e\rangle\langle e\vert +\vert
g\rangle \langle g\vert \right) +\lambda (a+a^{\dag
})\left(\vert e\rangle\langle g\vert +\vert
g\rangle \langle e\vert \right) $ \textcolor{black}{in} the total Hamiltonian $H(t)$ is the quantum Rabi Hamiltonian~\cite{Rabi1936}. The total 
Hamiltonian $H(t) $ can hence be written as $H(t)
=H_{\rm R}+\omega _{b}\vert b\rangle\langle b\vert
+\omega _{c}a^{\dag }a\vert b\rangle \langle b\vert
+\sum_{l=1}^{2}\left[ \Omega _{l}\left( t\right) \cos \left( \omega
_{l}t\right) \right] \left( \vert b\rangle \langle
g\vert +\vert g\rangle \langle b\vert \right) $.
In terms of the eigenstate $\left\vert \varepsilon_{n} \right\rangle $ ($n=0,1,\dotsb$) of \textcolor{black}{the QRM}, the first three terms in \textcolor{black}{$H(t)$} can be written in the
diagonal form:
\begin{equation}
	H_{0}=\sum_{n=0}^{\infty }\left[ \varepsilon _{n}\left\vert \varepsilon
	_{n}\right\rangle \left\langle \varepsilon _{n}\right\vert +\left( \omega
	_{b}+n\omega _{c}\right) \left\vert b,n\right\rangle \left\langle
	b,n\right\vert \right].  \label{Hamil_diag}
\end{equation}
The eigenstate $\vert \varepsilon _{n}\rangle $ can be expressed as $%
\vert \varepsilon _{n}\rangle =\sum_{m=0}^{\infty }\left(
C_{n,m}\vert g,m\rangle +D_{n,m}\vert e,m\rangle
\right) $ in terms of the uncoupled atomic and cavity states with real probability
amplitudes $C_{n,m}=\left\langle \varepsilon
_{n}|g,m\right\rangle $ and $D_{n,m}=\left\langle \varepsilon
_{n}|e,m\right\rangle $, which can be numerically obtained. Meanwhile, in terms of the
eigenstate $\left\vert\varepsilon_{n}\right\rangle$, the last term in \textcolor{black}{$H(t)$} can be written as 
\begin{equation}
	H_{D}\left( t\right) =\sum_{n,m=0}^{\infty }\sum_{l=1}^{2} \left[ \Omega
	_{l}\left( t\right) \cos \left( \omega _{l}t\right) 
	C_{n,m}\left\vert \varepsilon _{n}\right\rangle \left\langle b,m\right\vert +%
	\text{H.c.}\right].  \label{Hamil_d}
\end{equation}

In the rotating frame defined by the unitary operator $U(t)=%
\mathrm{exp}\left( -\text{i}H_{0}t\right) $, the system Hamiltonian becomes
\begin{equation}
	H_{I}\left( t\right) =\sum_{n,m=0}^{\infty }\sum_{l=1}^{2}\sum_{p=\pm 1}
	\left[ \Omega _{l,n,m}\left( t\right) e^{\text{i}\Delta _{n,m,p,l}t}
	\vert \varepsilon _{n}\rangle \langle b,m\vert +\text{%
		H.c.}\right],  \label{Hamil_I}
\end{equation}%
where the effective coupling strength $\Omega_{l,n,m}\left(t\right)$ and
the detuning $\Delta _{n,m,p,l}$ are 
\begin{eqnarray}
	\Omega _{l,n,m}\left(t\right) &=&\frac{C_{n,m}}{2}\Omega _{l}\left(t\right),%
	\hspace{0.5cm}(l=1,2),  \notag \\
	\Delta _{n,m,p,l} &=&\varepsilon _{n}-\omega _{b}-m\omega _{c}+p\omega _{l}.
\end{eqnarray}
Here the effective coupling strength is modified by the probability amplitudes $C_{n,m}$ ($n,m\ge0$ being integer) \textcolor{black}{of the eigenstates}. 
It is worth noting that the total number of excitations
in the QRM is not a conserved quantity due to the existence of the
counter-rotating terms, but the QRM possesses a parity (or $\mathbb{Z}_{2}$)
symmetry which shows that the system is integrable~\cite{Braak2011}. 
It can be shown that the ground \textcolor{black}{(first excited)} state of $H_{\rm R}$ only contains states with even \textcolor{black}{(odd)} number of excitations, and can be expanded as \textcolor{black}{$\vert\varepsilon_{M}\rangle=\sum^{\mathbb{\infty}}_{m=0}\left(%
	C_{M,2m+M}\left\vert g,2m\!+\!M\right\rangle+D_{M,2m+1-M}\left\vert e,2m\!+\!1\!-\!M\right\rangle
	\right)$ with $M=0$ \textcolor{black}{($1$)}}. Hence in the ground \textcolor{black}{(first excited)} state, the amplitudes of \textcolor{black}{the} states with odd \textcolor{black}{(even)} number of photons and the atom being in the state $\left\vert g\right\rangle$ satisfy $C_{0,2m+1}=0$ \textcolor{black}{($C_{1,2m}=0$)} due to the parity
symmetry~\cite{Braak2011}. The photons in the state $\vert\varepsilon_{0}\rangle$ are bounded (or virtual), and cannot be detected directly. To see the dependence of the virtual photon amplitudes on the coupling strength $\lambda$, we plot the
coefficient $C_{0,2m}$ ($m=0,1,2,3$) as a function of the
ratio $\lambda/\omega_{c}$ in Figure~\ref{fig_7_virtula_photon_Am}(a). 
When the coupling strength $\lambda/\omega_{c}$ is small, i.e., $\lambda/\omega_{c}\ll 0.1$, the Hamiltonian $H_{\rm R}$ is reduced to the Jaynes-Cummings (JC) Hamiltonian under the RWA. In this
regime, the ground state $\vert\varepsilon_{0}\rangle$ is mainly composed of $\left\vert g,0\right\rangle$ with zero excitation, and the population transfer from the state $\left\vert b,0\right\rangle$ to $\left\vert b,2m\right\rangle$ ($m>0$) will not occur due to the small amplitudes of the states $\left\vert g,2m\right\rangle$ ($m>0$). \textcolor{black}{However, with the increase of the coupling strength  $\lambda$,  the virtual-photon coefficient $C_{0,2m}$ ($m=1,2,\dotsb$) becomes significant which supports the population transfer from the state $\vert b,0\rangle$ to $\vert b,2m\rangle$.}  \textcolor{black}{It is worth noting that the population transfer from the state $\vert b,1\rangle$ to $\vert b,2m\!+\!1\rangle$ through the first excited state $\vert\varepsilon_{1}\rangle$  can also occur \textcolor{black}{when the amplitude} $C_{1,2m+1}$ ($m=1,2,\dotsb$) of the state $\vert g,2m\!+\!1\rangle$ in the first excited state $\vert \varepsilon_{1}\rangle$ \textcolor{black}{becomes significant for $\lambda/\omega_{c}\gtrsim 1$}, as shown in Figure~\ref{fig_7_virtula_photon_Am}(b). }
Therefore, in order to generate multiple photons, the atom-cavity coupling \textcolor{black}{strength} is required to be in the ultrastrong or deep-strong coupling regime, where \textcolor{black}{$C_{M,2m+M}$} ($m>0$) \textcolor{black}{can be} significant.

\section{STIRAP generation of multiple photons}~\label{three_level_eff_haml}
In this section, we derive an effective Hamiltonian of the above system under the $2m$-photon ($m$ being positive integer) resonance condition~\cite{Huang2014} 
\begin{equation}
	\omega _{1}-\omega _{2}=2m\omega _{c}, \label{ren_condition}
\end{equation}%
and elucidate the physical mechanism of \textcolor{black}{the} coherent population transfer from
the initial state \textcolor{black}{$\left\vert b,M\right\rangle $} to the final state \textcolor{black}{$\left\vert
	b,2m\!+\!M\right\rangle $} based on the STIRAP technique. We will then discuss
the generation of \textcolor{black}{multiple} photons in detail.

\subsection{Effective Hamiltonian and photon generation}
Assume that the system is in the initial state \textcolor{black}{$\left\vert b,M\right\rangle$} with the cavity field in \textcolor{black}{the Fock} \textcolor{black}{state $\vert M\rangle$} and the atom in the lowest state $\left\vert b\right\rangle$.
Let the driving frequencies be near resonance with both the transition frequency \textcolor{black}{$\varepsilon_{n}-\omega_b-M\omega_{c}$} \textcolor{black}{with $M$ photons} and the \textcolor{black}{(2$m$+$M$)}-photon emission frequency \textcolor{black}{$\varepsilon_{n}-\omega_b-(2m\!+\! M)\omega_c$}, respectively, i.e., \textcolor{black}{$\Delta _{n,M,-1,1}, \Delta_{n,2m+M,-1,2}\ll 2m\omega _{c}$}, and the \textcolor{black}{other eigenstates $\vert\varepsilon_{n'}\rangle$ ($n'\!\neq\! n$)} are far detuned from the driving frequencies so that they can be neglected from this scheme. 
Under the 2$m$-photon resonance condition (\ref{ren_condition}), the Hamiltonian (\ref{Hamil_I}) can be reduced to 
\textcolor{black}{\begin{eqnarray}
		\tilde{H}_{I}\left( t\right) &=&\Omega _{1,n,M}\left( t\right) e^{\text{i}\Delta
			_{n,M,-1,1}t}\left\vert \varepsilon _{n}\right\rangle \left\langle
		b,M\right\vert  \notag \\
		&&+\Omega _{2,n,2m+M}\left( t\right) e^{\text{i}\Delta _{n,2m+M,-1,2}t}\left\vert
		\varepsilon _{n}\right\rangle \left\langle b,2m\!+\!M\right\vert +\text{H.c.},\notag\\
		\label{H_titrd}
\end{eqnarray}}\textcolor{black}{with $M=n$ mod 2 (\textcolor{black}{i.e.,} $M=0$ or 1 corresponds to the even-photon or odd-photon bundles case, \textcolor{black}{respectively}). We find that eq.~(\ref{H_titrd}) connects the states $\left\vert b,2m\!+\!M\right\rangle$ with $2m$+$M$ photons to the eigenstates $\left\vert \varepsilon _{n}\right\rangle $  of the QRM}. Here we have ignored the fast oscillating terms under the RWA by the condition $\vert\Omega_{l,n,m}(t)/
\Delta_{n,m,+1,l}\vert \ll 1$ and \textcolor{black}{$\vert \Omega _{l,n',m}(t)/\Delta_{n',m,-1,l}\vert \ll 1   \hspace{0.1cm}$ ($n'\!\neq\! n$, $l=1,2$).} 
The system can then be reduced to an effective three-level system, as shown in Figure~\ref{Fig1_model}(b).

With the $2m$-photon resonance condition (\ref{ren_condition}), \textcolor{black}{$\Delta
	_{n,M,-1,1}=\Delta _{n,2m+M,-1,2}\equiv\Delta $}. In a rotating frame with respect to  \textcolor{black}{$\tilde{H}_{0}=-\Delta\vert\varepsilon_{n}\rangle\langle\varepsilon_{n}\vert$},
we obtain the effective Hamiltonian for this system: 
\textcolor{black}{\begin{eqnarray}
		H_{\mathrm{eff}}^{\left( 2m+M\right) } &=&\Delta\left\vert \varepsilon _{n}\right\rangle \left\langle \varepsilon _{n}\right\vert
		+[\Omega _{1,n,M}\left( t\right) \left\vert \varepsilon _{n}\right\rangle
		\left\langle b,M\right\vert   \notag \\
		&&+\Omega _{2,n,2m\!+\!M}\left( t\right) \left\vert \varepsilon _{n}\right\rangle
		\left\langle b,2m+M\right\vert +\text{H.c.}].  \label{Heffff}
\end{eqnarray}}This Hamiltonian describes a $\Lambda$-type three-level system, where the effective coupling strength \textcolor{black}{$\Omega_{1,n,M}\left(t\right)$ [$\Omega_{2,n,2m+M}\left(t\right)$] depends on the  coefficient $C_{n,M}$ ($C_{n,2m+M}$)}. The \textcolor{black}{effective} coupling strength \textcolor{black}{$\Omega_{1,n,M}\left(t\right)$ [$\Omega_{2,n,2m+M}\left(t\right)$]} can be tuned by choosing appropriate coupling strength $\lambda$, which can strongly affect the coefficients \textcolor{black}{$C_{n,M}$ and $C_{n,2m+M}$} according to Figure~\ref{fig_7_virtula_photon_Am}. The transfer from the state \textcolor{black}{$\left\vert b,M\right\rangle$} with the cavity in the  state \textcolor{black}{$\vert M\rangle$} to the state \textcolor{black}{$\left\vert b,2m\!+\!M\right\rangle$} with \textcolor{black}{$2m\!+\!M$} cavity photons can be achieved through these two \textcolor{black}{effective} couplings, which is the key mechanism for our
scheme to generate multiple photons. 

In our scheme, the coupling strengths \textcolor{black}{$\Omega_{1,n,M}\left(t\right)$ and $\Omega_{2,n,2m+M}\left(t\right)$} in eq.~(\ref{Heffff}) are time dependent. We derive the instantaneous
eigenstates of the effective Hamiltonian (\ref{Heffff}) at time $t$ as follows
\textcolor{black}{\begin{subequations}
		\begin{align}
			\left\vert \psi _{0}^{\left( 2m+M\right) }\left( t\right) \right\rangle
			& =\cos \left[ \theta _{2m+M}\left( t\right) \right] \left\vert
			b,M\right\rangle   \notag \\
			& \hspace{0.3cm} -\sin \left[ \theta _{2m+M}\left( t\right) \right] \left\vert
			b,2m\!+\!M\right\rangle ,  \label{dark_state} \\
			\left\vert \psi _{+}^{\left( 2m+M\right) }\left( t\right) \right\rangle
			& =\sin \left[ \varphi _{2m+M}\left( t\right) \right] \{\sin \left[ \theta
			_{2m+M}\left( t\right) \right] \left\vert b,M\right\rangle   \notag \\
			& \hspace{0.3cm}+\cos \left[ \theta _{2m+M}\left( t\right) \right]
			\left\vert b,2m\!+M\!\right\rangle \}  \notag \\
			& \hspace{0.3cm} +\cos \left[ \varphi _{2m+M}\left( t\right) \right] \left\vert \varepsilon
			_{n}\right\rangle , \\
			\left\vert \psi _{-}^{\left( 2m+M\right) }\left( t\right) \right\rangle
			& =\cos \left[ \varphi _{2m+M}\left( t\right) \right] \{\sin \left[ \theta
			_{2m+M}\left( t\right) \right] \left\vert b,M\right\rangle   \notag \\
			& \hspace{0.3cm}+\cos \left[ \theta _{2m+M}\left( t\right) \right]
			\left\vert b,2m\!+\!M\right\rangle \}  \notag \\
			& \hspace{0.3cm} -\sin \left[ \varphi _{2m+M}\left( t\right) \right] \left\vert \varepsilon
			_{n}\right\rangle ,
		\end{align}
\end{subequations}}and the corresponding instantaneous eigenvalues are $\lambda _{0}=0,$ \textcolor{black}{$\lambda
	_{+}=\tilde{\Omega}_{2m+M}\left( t\right) \cot [\varphi _{2m+M}\left( t\right)] $,
	and $\lambda _{-}=-\tilde{\Omega}_{2m+M}\left( t\right) \tan [\varphi
	_{2m+M}\left( t\right)]$,} where 
\textcolor{black}{\begin{subequations}
		\begin{align}
			\theta _{2m+M}\left( t\right)  &=\arctan\left[\frac{\eta _{2m+M}\Omega _{1}\left(
				t\right) }{\Omega _{2}\left( t\right) }\right],   \\
			\varphi _{2m+M}\left( t\right)  &=\arctan\left[\frac{\tilde{\Omega}_{2m+M}\left(
				t\right) }{\frac{\Delta}{2}+\sqrt{\frac{\Delta ^{2}}{4}+\tilde{\Omega}
					_{2m+M}^{2}\left( t\right) }}\right],
		\end{align}
\end{subequations}}with
\textcolor{black}{\begin{equation}
		\tilde{\Omega}_{2m+M}\left( t\right) =\frac{|C_{n,2m+M}|}{2}\sqrt{\eta
			_{2m+M}^{2}\Omega _{1}^{2}\left( t\right) +\Omega _{2}^{2}\left( t\right) }
\end{equation}}and \textcolor{black}{$\eta _{2m+M}=|C_{n,M}/C_{n,2m+M}|$.}
The eigenstate \textcolor{black}{$\vert \psi _{0}^{\left( 2m+M\right) }\left( t\right) \rangle$} in eq.~(\ref{dark_state}) with eigenvalue $\lambda _{0}=0$ is a dark state, which does not include the state 
\textcolor{black}{$\left\vert \varepsilon	_{n}\right\rangle$} as its component. Instead, the dark state is a coherent superposition of the \textcolor{black}{$M$-photon state} and the \textcolor{black}{($2m+M$)}-photon state of the cavity with the atom in the lowest level $\left\vert b\right\rangle $. With the system initially prepared in the dark state \textcolor{black}{$
	\vert \psi _{0}^{\left( 2m+M\right) }\left( t\right)\rangle $} and the effective coupling strengths tuned adiabatically under the condition~\cite{ Bergmann1998,Vitanov2017,Liu2022}: \textcolor{black}{$\vert\dot{\theta}_{2m+M}\left(t\right)\vert\ll\vert\lambda_{\pm}-\lambda_{0}\vert$}, which leads to
\textcolor{black}{\begin{equation}
		\left\vert \dot{\theta}_{2m+M}\left( t\right) \right\vert \ll \left\vert \frac{%
			\Delta }{2}\pm \sqrt{\frac{\Delta ^{2}}{4}+\tilde{\Omega}_{2m+M}^{2}\left(
			t\right) }\right\vert ,  \label{evo_con}
\end{equation}}the system will remain in the dark state \textcolor{black}{$\vert\psi^{\left(2m+M\right)}_{0}\left(t%
	\right)\rangle$} at an arbitrary time $t$ during the evolution. Thus, by adjusting the driving amplitudes appropriately, the system state can be converted from an initial dark state to a desired dark state at the end of the evolution using the STIRAP technique. 
In our approach, the initial state at $t=0$ is prepared in the dark state  \textcolor{black}{$\vert\psi^{(2m+M)}_{0}\left(0\right)\rangle=\vert b,M\rangle$} for \textcolor{black}{$\theta_{2m+M}\left( 0\right)=0$}, and at the final time $t$,  \textcolor{black}{$\theta_{2m+M}\left( t\right)=\pi/2$}, which corresponds to the dark state \textcolor{black}{$\left\vert b,2m+M\right\rangle$}. By increasing \textcolor{black}{$\theta_{2m}\left( t\right)$} adiabatically under the condition (\ref{evo_con}), we can hence convert the state \textcolor{black}{$\vert b,M\rangle$} to the multi-photon state \textcolor{black}{$\vert b,2m+M\rangle$}. 
The merit of the STIRAP
\begin{figure}[tbp]
	\center
	\includegraphics[width=8.3cm]{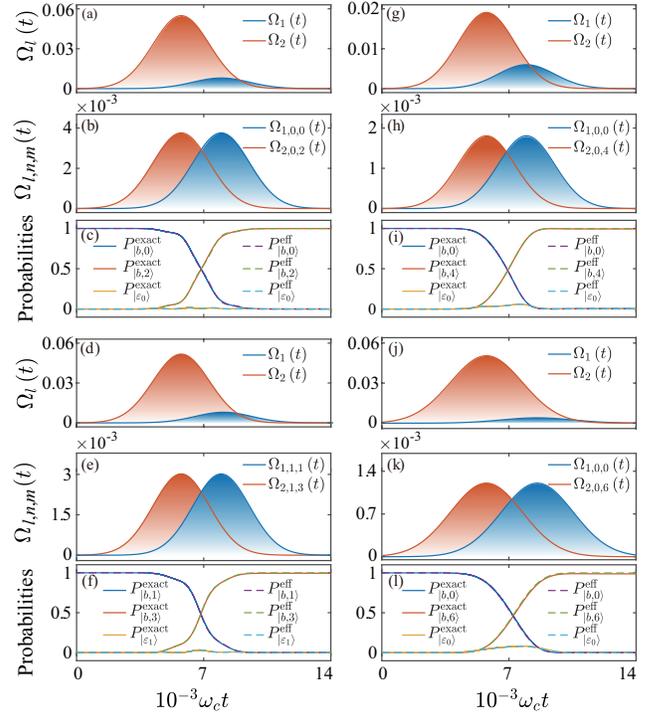}
	\caption{(Color online) \textcolor{black}{Dynamics of multi-photon generation (two photons (a)-(c), three photons (d)-(f),  four photons (g)-(i) and six photons (j)-(l)) with STIRAP. (a), (d), (g), (j)  The	amplitudes of the Gaussian pulses in one period for two-, three-, four- and six-photon 
			generation, respectively. (b), (e), (h), (k) The corresponding effective coupling strengths 
			$\Omega _{1,0,0}\left( t\right)$, $\Omega _{1,1,1}\left( t\right)$, $\Omega _{2,1,3}\left( t\right)$,and $\Omega _{2,0,2m}\left( t\right)$ ($m$=1,2,3) based on eq.~(\protect\ref{Heffff}). (c), (f), (i), (l) The time dependence of the probabilities of  the states $\left\vert b,2m\right\rangle$ ($m$%
			=0,1,2,3),  $\left\vert b,2m+1\right\rangle$ ($m$%
			=0,1) and $\vert\protect\varepsilon_{n}\rangle$ ($n$=0,1) under the driving
			pulses given in \textcolor{black}{panels} (a), (d), (g) and (j) respectively. The result $(P^{\mathrm{eff}}_{\vert\cdot\rangle})$ from the effective three-level system (dotted curves) agrees well with the result $(P^{\mathrm{exact}}_{\vert\cdot\rangle})$ from the exact Hamiltonian (solid curves). The parameters are (a)-(c) $\Omega
			_{2}/\Omega _{1}=6.8538$, (d)-(f) $\Omega_{2}/\Omega _{1}=6.4641$, 	(g)-(i) $\Omega _{1}/\protect\omega %
			_{c}=0.006$, $\Omega _{2}/\Omega _{1}=3.1814$, $\protect\omega _{c}t_{1}=7960$, $	\omega _{c}T=2200$, and (j)-(l) $\Omega _{1}/\protect\omega 
			_{c}=0.004$, $\Omega _{2}/\Omega _{1}=12.6179$, $\protect\omega _{c}t_{1}=8560$,  $	\omega _{c}T=2800$. The common parameters are $\protect\lambda /\protect\omega _{c}=0.6$, $\protect\omega _{b}/%
			\protect\omega _{c}=-6$, $\Omega _{1}/\protect\omega _{c}=0.008$, $\protect\omega _{c}t_{1}=7960$, $	\omega _{c}T=2200$ in \textcolor{black}{panels} (a)-(f),
			and $\protect\lambda /\protect\omega _{c}=1.2$, $\protect\omega _{b} /
			\protect\omega _{c}=-10$ in \textcolor{black}{panels} (g)-(l). Other parameters are $%
			\protect\omega _{e}-\protect\omega _{g}=\protect\omega _{c}$, $\Delta=0$, $\protect\omega _{c}t_{2}=5760$ and $\omega _{c}T_{1}=84000$.}}
\label{Fig2_two_four_photons_v1}
\end{figure}
\noindent technique is that it only involves the dark state (not the state \textcolor{black}{$\vert\varepsilon_{n}\rangle$}), which will not decay to the lowest atomic level $\left \vert b\right \rangle$ via spontaneous emission. Moreover, as $\vert b\rangle$ is not coupled  to the cavity mode, the photons are emitted only through the STIRAP process. The process is thus fully deterministic via external control fields.  

In the following, \textcolor{black}{we will discuss the generation of multiple photons in detail  for the detuning $\Delta =0$,} \textcolor{black}{and demonstrate the physical mechanism of the generation of  multi-photon \textcolor{black}{state} via the STIRAP technique based on the dark state~(\ref{dark_state}).}
\subsection{Two-photon generation}
We first choose \textcolor{black}{$n=0$, $M=0$ \text{and} $m$=1} in the Hamiltonian~(\ref{Heffff}) for the generation of two photons, which requires $\omega_1-\omega_2=2\omega_c$.
At $\Delta=0$, the effective Hamiltonian (\ref{Heffff}) becomes
\begin{equation}
H_{\mathrm{eff}}^{\left( 2\right) }\left( t\right) =\Omega _{1,0,0}\left(
t\right) \left\vert \varepsilon _{0}\right\rangle \left\langle
b,0\right\vert +\Omega _{2,0,2}\left( t\right) \left\vert \varepsilon
_{0}\right\rangle \left\langle b,2\right\vert +\text{H.c.},
\label{Hamil_eff_2}
\end{equation}%
and the corresponding dark state (\ref{dark_state}) becomes 
\textcolor{black}{\begin{equation}
	\left\vert \psi _{0}^{\left( 2\right) }\left( t\right) \right\rangle =\cos
	[\theta _{2}\left( t\right)] \left\vert b,0\right\rangle -\sin [\theta
	_{2}\left( t\right)] \left\vert b,2\right\rangle
\end{equation}}with $\theta _{2}\left( t\right) =\arctan \left[ \eta
_{2}\Omega _{1}\left( t\right) /\Omega _{2}\left( t\right) \right]$.

To generate two photons with STIRAP, the system is required to adiabatically follow the dark state $\vert \psi _{0}^{\left( 2\right) }\left( t\right) \rangle $ during the evolution~\cite{Bergmann1998,Vitanov2017}. 
Let the initial state at time $t=0$ be the dark state $\vert \psi _{0}^{\left( 2\right) }\left(0\right)\rangle=\vert b,0\rangle $ for $\theta_{2}\left(0\right)=0$, which requires $
\Omega _{1}\left( 0\right) /\Omega _{2}\left( 0\right) \rightarrow 0$. 
We then adiabatically change  $\theta _{2}\left(t\right)$ to reach  $\theta _{2}\left(t\right)= \pi /2$, which corresponds to $\Omega _{1}\left(
t\right) /\Omega _{2}\left( t\right) \rightarrow \infty$. At time $t$, the dark state is  
$\vert \psi _{0}^{\left( 2\right) }\left( t\right)\rangle
= \left\vert b,2\right\rangle$, which is the atomic state $\left\vert b\right\rangle$ plus two cavity photons. 
Note that the efficient implementation of the STIRAP process requires that the two pulses $\Omega _{1}\left(
t\right)$ and $\Omega _{2}\left( t\right)$ have significant overlap in time, as shown in Figure~\ref{Fig2_two_four_photons_v1}(a). The effective coupling strengths $\Omega _{1,0,0}\left( t\right) $ and $\Omega _{2,0,2}\left( t\right) $ are plotted in Figure~\ref{Fig2_two_four_photons_v1}(b), where the maximum values of the two couplings are equal to each other. It is worth mentioning that the pulse $\Omega _{2}\left( t\right)$ is applied before $\Omega _{1}\left( t\right)$, which is counter-intuitive but typical in STIRAP. 
To demonstrate the photon generation in this process, we plot the probabilities of the states $\left\vert b,0\right\rangle$, $\left\vert
b,2\right\rangle$, and $\left\vert \varepsilon _{0}\right\rangle $ in Figure~\ref{Fig2_two_four_photons_v1}(c), from simulations of both the effective Hamiltonian and the exact Hamiltonian. Our result shows that the population in $\vert b,0\rangle$ can be almost completely transferred to $\vert b,2\rangle$ with its final probability reaching 1, and the probability of $\vert\varepsilon_{0}\rangle$ \textcolor{black}{being on the order of $10^{-4}$}
at the end of the STIRAP process. With the above driving parameters, the population transfer from $\vert b,2\rangle$ to $\vert b,4\rangle$ is negligible with the probability of $\vert b,4\rangle$ on the order of $10^{-4}$. This is due to the large detuning between the driving frequencies and their corresponding transition frequencies. We want to emphasize that the result [$P^{\mathrm{exact}}_{\vert\cdot\rangle}$ in Figure~\ref{Fig2_two_four_photons_v1}(c)] from simulating the exact total Hamiltonian~(\ref{Hamil_I}) agrees well with the result [$P^{\mathrm{eff}}_{\vert\cdot\rangle}$ in Figure~\ref{Fig2_two_four_photons_v1}(c)] from simulating the effective three-level Hamiltonian (\ref{Hamil_eff_2}), which verifies that the three-level approximation is valid.   

\textcolor{black}{\subsection{Four-photon and six-photon generation}}
For \textcolor{black}{$n=0$, $M=0$, $m=2$ and $\Delta=0$}, the effective Hamiltonian (\ref{Heffff}) becomes
\begin{equation}
H_{\mathrm{eff}}^{\left( 4\right) }\left( t\right) =\Omega _{1,0,0}\left(
t\right) \left\vert \varepsilon _{0}\right\rangle \left\langle
b,0\right\vert +\Omega _{2,0,4}\left( t\right) \left\vert \varepsilon
_{0}\right\rangle \left\langle b,4\right\vert +\text{H.c.},
\label{Hamil_eff_4}
\end{equation}%
and the corresponding dark state (\ref{dark_state}) has the form 
\textcolor{black}{\begin{equation}
	\left\vert \psi _{0}^{\left( 4\right) }\left( t\right) \right\rangle =\cos
	[\theta _{4}\left( t\right)] \left\vert b,0\right\rangle -\sin [\theta
	_{4}\left( t\right)] \left\vert b,4\right\rangle ,
\end{equation}}with $\theta _{4}\left( t\right) =\arctan \left[ \eta
_{4}\Omega _{1}\left( t\right) /\Omega _{2}\left( t\right) \right]$. The physical process here is similar to the process of the two-photon generation, 
and the efficient generation of four photons also requires an appropriate overlap between $\Omega _{1}\left(
t\right)$ and $\Omega _{2}\left( t\right)$, as shown in Figure~\ref{Fig2_two_four_photons_v1}(g). The  corresponding effective coupling strengths $\Omega _{1,0,0}\left( t\right) $
and $\Omega _{2,0,4}\left( t\right) $ are plotted in Figure~\ref{Fig2_two_four_photons_v1}(h), which also shows that the peak values of the couplings are equal to each other. The probabilities of the states $\left\vert b,0\right\rangle$, $\left\vert
b,4\right\rangle$ and $\left\vert \varepsilon _{0}\right\rangle $ are plotted in Figure~\ref{Fig2_two_four_photons_v1}(i), which shows that the population in the state $\vert b,0\rangle$ can almost be completely transferred to the state $\vert b,4\rangle$ with the probability of $\vert b,4\rangle$ approaching 1, and the probability of $\vert\varepsilon_{0}\rangle$ \textcolor{black}{being on the order of $10^{-4}$}
at the end of the STIRAP. The \textcolor{black}{population} transfer from the state $\vert b,4\rangle$ to the state $\vert b,8\rangle$ is negligible with the probability of $\vert b,8\rangle$ on the order of $10^{-6}$ due to the large detuning in the corresponding transition. The simulation data based on the exact Hamiltonian and the effective Hamiltonian match also well in this case. This result shows that the
generation of four photons can be implemented in our scheme.

\textcolor{black}{Similarly,  the generation of  six photons \textcolor{black}{can be easily achieved} with the same mechanism due to the  significant amplitude coefficient $C_{0,6}$, as shown in Figure~\ref{fig_7_virtula_photon_Am}(a). 
It is worth noting that the second \textcolor{black}{($\lambda/\omega_{c}<0.43$) or third ($\lambda/\omega_{c}>0.43$)} excited state $\vert \varepsilon_{2}\rangle$ only include the even number of excitations due to its even parity with the current parameter \textcolor{black}{$\omega_{c}\!=\!\omega_{e}\!-\!\omega_{g}$},  and can be expanded as $\vert\varepsilon_{2}\rangle=\sum^{\mathbb{\infty}}_{m=0}\left(
C_{2,2m}\left\vert g,2m\right\rangle+D_{2,2m+1}\left\vert e,2m\!+\!1\right\rangle\right)$.
In Figure~\ref{fig_7_virtula_photon_Am}(a), we find that the coefficient $C_{2,6}$ of $\vert g,6\rangle$ in the eigenstate $\vert\varepsilon_{2}\rangle$ of the QRM \textcolor{black}{is} larger than $C_{0,6}$. Thus the scheme  of six-photon generation via the second excited state $\vert\varepsilon_{2}\rangle$ is feasible. Here we choose the ground state $\vert \varepsilon_{0}\rangle$ of the QRM to generate six photons as an example, and the effective Hamiltonian can be written as
\begin{equation}
	H_{\mathrm{eff}}^{\left( 6\right) }\left( t\right) =\Omega _{1,0,0}\left(
	t\right) \left\vert \varepsilon _{0}\right\rangle \left\langle
	b,0\right\vert +\Omega _{2,0,6}\left( t\right) \left\vert \varepsilon
	_{0}\right\rangle \left\langle b,6\right\vert +\text{H.c.},
	\label{Hamil_eff_6}
\end{equation}
and the corresponding dark state (\ref{dark_state}) has the form 
\textcolor{black}{\begin{equation}
		\left\vert \psi _{0}^{\left( 6\right) }\left( t\right) \right\rangle =\cos
		[\theta _{6}\left( t\right)] \left\vert b,0\right\rangle -\sin [\theta
		_{6}\left( t\right)] \left\vert b,6\right\rangle ,
\end{equation}}with $\theta _{6}\left( t\right) =\arctan \left[ \eta
_{6}\Omega _{1}\left( t\right) /\Omega _{2}\left( t\right) \right]$. The physical process of six-photon generation is similar to the process of the four-photon generation under the appropriate parameters, as shown in Figure~\ref{Fig2_two_four_photons_v1}(j-l). The \textcolor{black}{population} in the state $\vert b,0\rangle$ can almost be completely transferred to the state $\vert b,6\rangle$ with the probability approaching  1 after the STIRAP process, and other population transfer \textcolor{black}{processes} are negligible due to \textcolor{black}{large detuning and} \textcolor{black}{tiny} probabilities. It is worth noting that the simulation data based on the exact Hamiltonian and the effective Hamiltonian match also well in this situation, which shows that the generation of six photons can be implemented in our scheme.}

\textcolor{black}{\subsection{Three-photon generation}}
\textcolor{black}{To generate three photons, we choose the first excited state \textcolor{black}{$\vert\varepsilon_{1}\rangle$} which only includes odd-photon \textcolor{black}{component} $\vert g,2m+1\rangle$ ($m=0,1,\dotsb$) \textcolor{black}{for state $\vert g\rangle$ and even-photon component $\vert e,2m\rangle$ for state $\vert e\rangle$} due to the parity symmetry. Similarly, we choose $n=1$, $M=1$, $m=1$ and $\Delta=0$ under the two-photon resonance condition, \textcolor{black}{thus} the effective Hamiltonian (\ref{Heffff}) becomes
\begin{equation}
	H_{\mathrm{eff}}^{\left( 3\right) }\left( t\right) =\Omega _{1,1,1}\left(
	t\right) \left\vert \varepsilon _{1}\right\rangle \left\langle
	b,1\right\vert +\Omega _{2,1,3}\left( t\right) \left\vert \varepsilon
	_{1}\right\rangle \left\langle b,3\right\vert +\text{H.c.},
	\label{Hamil_eff_3}
\end{equation}%
and the corresponding dark state (\ref{dark_state}) has the form 
\textcolor{black}{\begin{equation}
		\left\vert \psi _{0}^{\left( 3\right) }\left( t\right) \right\rangle =\cos
		[\theta _{3}\left( t\right)] \left\vert b,1\right\rangle -\sin [\theta
		_{3}\left( t\right)] \left\vert b,3\right\rangle ,
\end{equation}}with $\theta _{3}\left( t\right) =\arctan \left[ \eta
_{3}\Omega _{1}\left( t\right) /\Omega _{2}\left( t\right) \right]$. The physical mechanism of three-photon generation is similar to two-photon case. To generate three photons with STIRAP, we assume that the initial system is in state $\vert b,1\rangle$, and the system is required to adiabatically follow the dark state $\psi^{(3)}_{0}\left(t\right)$ during the evolution~\cite{Bergmann1998,Vitanov2017}. Similarly, a appropriate overlap between $\Omega_{1}\left(t\right)$ and $\Omega_{2}\left(t\right)$  needs to be satisfied for the efficient generation of three photons, as shown in Figure~\ref{Fig2_two_four_photons_v1}(d),  and the effective coupling strengths $\Omega _{1,1,1}\left( t\right) $ and $\Omega _{2,1,3}\left( t\right) $ are plotted in Figure~\ref{Fig2_two_four_photons_v1}(e), where \textcolor{black}{the} maximum values are equal to each other. \textcolor{black}{Note} that the probabilities  of the states $\vert b,1\rangle$, $\vert b,3\rangle$ and $\vert \varepsilon_{1}\rangle$ are plotted in Figure~\ref{Fig2_two_four_photons_v1}(f), which demonstrates that the \textcolor{black}{population} in the state $\vert b,1\rangle$ is transferred adiabatically to the state $\vert b,3\rangle$ with the \textcolor{black}{probability} of $\vert b,3\rangle$ approaching 1, and the probability of $\vert \varepsilon_{1}\rangle$ being on the order of $10^{-4}$ at the end of the STIRAP. Under the driving pulses given in Figure~\ref{Fig2_two_four_photons_v1}(d), the population transfer from the state $\vert b,3\rangle$ to the state $\vert b,5\rangle$ is negligible with the probability of $\vert b,5\rangle$ on the order of $10^{-3}$ due to the large detuning in the corresponding transitions. The simulation data from the exact Hamiltonian and from the effective Hamiltonian \textcolor{black}{agree} well in this case. This result shows that the generation of three photons is feasible  by adjusting the driving frequencies of driving pulses to aim \textcolor{black}{at} the first excited state $\vert\varepsilon_{1}\rangle$ of the QRM.}

\
\section{Emission of multi-photon bundles~\label{dynamic}}
Photons generated in the above STIRAP process will be emitted to the cavity output via cavity dissipation. In this section, we study the multi-photon bundle emission using the following quantum
master equation~\cite{Ridolfo2012}, 
\begin{eqnarray}
\dot{\rho}\left( t\right) &=&-\text{i}\left[ H\left( t\right) ,\rho \left( t\right) %
\right]  \notag \\
&&+\sum_{u=a,ge,bg}\sum^{\infty}_{n,m>n}\Gamma _{u}^{m,n}\left\{ \mathcal{D}\left[
\left\vert \psi _{n}\right\rangle \left\langle \psi _{m}\right\vert \right]
\rho \left( t\right) \right\} ,  \label{master}
\end{eqnarray}%
with the superoperator $\mathcal{D}[ O] \rho( t)\!=\!O\rho ( t) O^{\dag }\!-\!O^{\dag }O\rho( t)/2 \!-\!\rho \left( t\right) O^{\dag }O/2$. Here, we assume that the system-bath coupling is weak with a zero-temperature Markovian bath~\cite{breuer2002}, and that the total Hamiltonian $H(t)$ including the driving
fields is
\begin{figure}[tbp]
\center
\includegraphics[width=8.3cm]{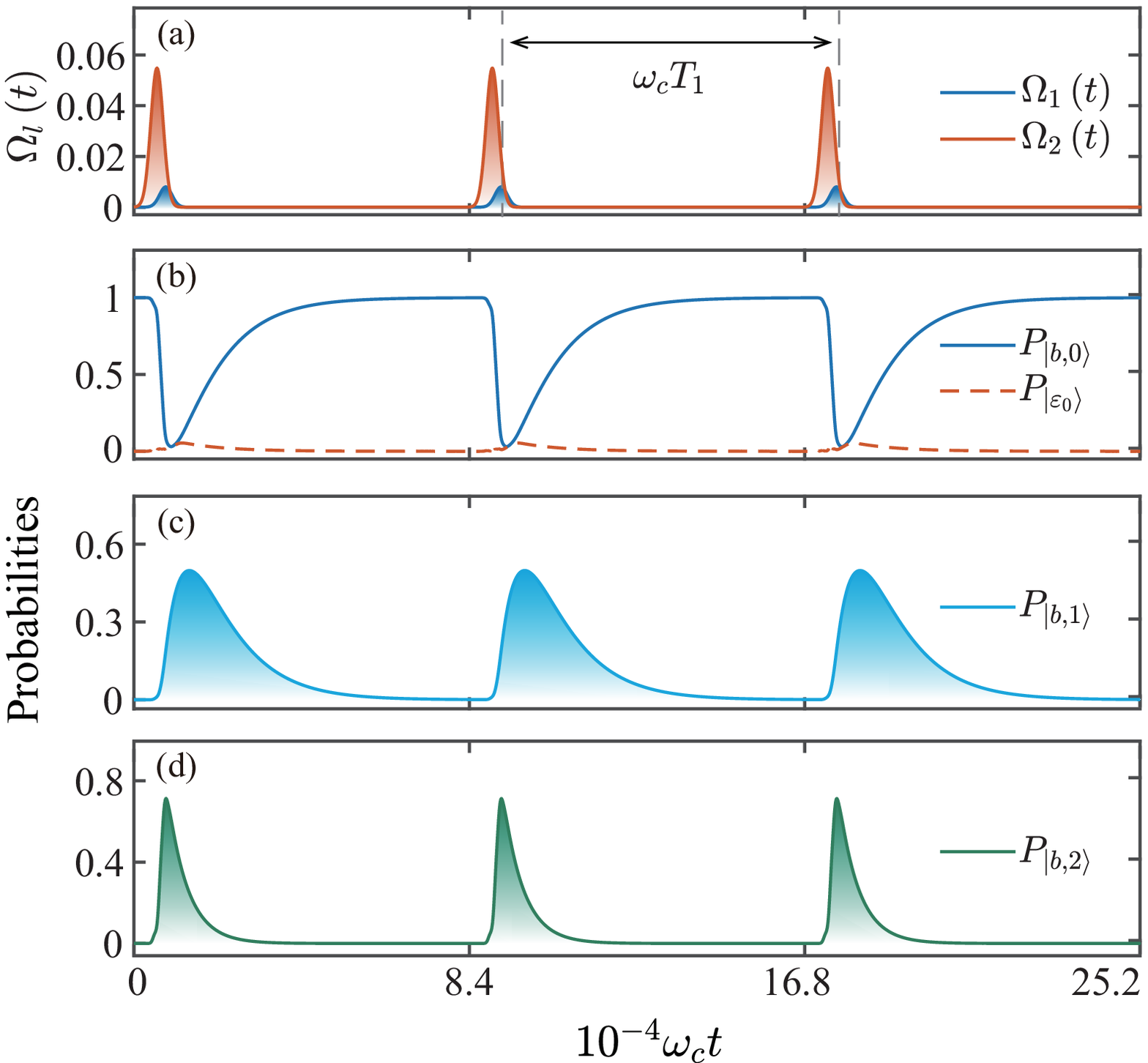}
\caption{(Color online) (a) The amplitudes $\Omega_{l}\left(t\right)$ ($l$%
	=1,2) of the Gaussian pulses as a function of the scaled time $ 10^{-4}
	\protect\omega_{c}t$. (b)-(d) The probabilities $%
	P_{\left\vert b,j\right\rangle}\left(t\right)$ ($j$=0,1,2) and $P_{\vert
		\protect\varepsilon_{0}\rangle}\left( t\right)$ vs. the scaled time. 
	The decay rates are $\protect\kappa_{a}/\protect\omega_{c} =\protect\kappa_{ge}/\protect\omega_{c}=\protect\kappa
	_{bg}/\protect\omega_{c}=0.0001$. Other parameters are the same as those in Figure~%
	\protect\ref{Fig2_two_four_photons_v1}(a)-(c).}
\label{Fig3_two_photons_diss}
\end{figure}
\noindent given by eq.~(\ref{H}). The state $\left\vert \psi _{n}\right\rangle $ is an eigenstate of the Hamitonian $H_{0}$ with eigenenergy $E_n$, i.e., $\left\{ \left\vert \psi _{n}\right\rangle \right\} =\left\{\left\vert b,0\right\rangle ,\left\vert b,1\right\rangle ,\left\vert
b,2\right\rangle ,...,\left\vert \varepsilon _{0}\right\rangle ,\left\vert
\varepsilon _{1}\right\rangle ,...\right\} $. The relaxation rate in the superoperator $\mathcal{D}[ O]$ is defined as
\begin{equation}
\Gamma _{u}^{m,n}=2\pi d_{u}\left( \Delta _{m,n}\right) \alpha
_{u}^{2}\left( \Delta _{m,n}\right) \left\vert C_{u}^{n,m}\right\vert ^{2}, 
\hspace{0.5cm}(u=a,\,ge,\,bg),
\end{equation}
which is determined by the spectral density $d_{u}\left( \Delta_{m,n}\right)$ of the bath modes, 
the system-bath coupling strength $\alpha _{u}\left( \Delta_{m,n}\right)$, and the transition matrix elements
\begin{eqnarray}
C_{a}^{n,m} &=&\left\langle \psi _{n}\right\vert (a+a^{\dag })\left\vert
\psi _{m}\right\rangle ,  \notag \\
C_{ge}^{n,m} &=&\left\langle \psi _{n}\right\vert (\left\vert g\right\rangle
\left\langle e\right\vert +\left\vert e\right\rangle \left\langle
g\right\vert )\left\vert \psi _{m}\right\rangle , \\
C_{bg}^{n,m} &=&\left\langle \psi _{n}\right\vert (\left\vert b\right\rangle
\left\langle g\right\vert +\left\vert g\right\rangle \left\langle
b\right\vert )\left\vert \psi _{m}\right\rangle.  \notag
\end{eqnarray}%
Here, $u=a$ denotes the bath for cavity damping, $u=ge,\, bg$ denote
the baths that induce the atomic decay from $\left\vert e\right\rangle $ to $%
\left\vert g\right\rangle $ and from $\left\vert g\right\rangle $ to $\left\vert
b\right\rangle $, respectively, and $\Delta _{m,n}=E_{m}-E_{n}$ is the
transition frequency between the states $\left\vert \psi _{m}\right\rangle $ and $%
\left\vert \psi _{n}\right\rangle $. Note that we have neglected the Lamb-shift terms in eq.~(\ref{master}). For simplicity of discussion, we assume that the spectral density $%
d_{u}\left( \Delta _{m,n}\right) $ and the system-bath coupling strength $%
\alpha _{u}\left( \Delta _{m,n}\right) $ be constant with the decay rate 
\begin{equation}
\kappa _{u}=2\pi d_{u}\left( \Delta _{m,n}\right) \alpha _{u}^{2}\left(
\Delta _{m,n}\right), \hspace{0.5cm}(u=a,\,ge,\,bg).
\end{equation}%
The relaxation coefficients are then $\Gamma
_{u}^{m,n}=\kappa _{u}\vert C_{u}^{n,m}\vert ^{2}.$

To investigate how the cavity photons are emitted to the cavity output, we
numerically calculate the probabilities $P_{\left\vert b,j\right\rangle
}\left( t\right) $ $\left( j=0,1,\dotsb\right) $ of the state $\left\vert b,j\right\rangle $ and $P_{\left\vert \varepsilon_{0}\right\rangle }\left( t\right) $ [$P_{\left\vert \varepsilon_{1}\right\rangle }\left( t\right) $]
\begin{figure}[tbp]
\center
\includegraphics[width=8.3cm]{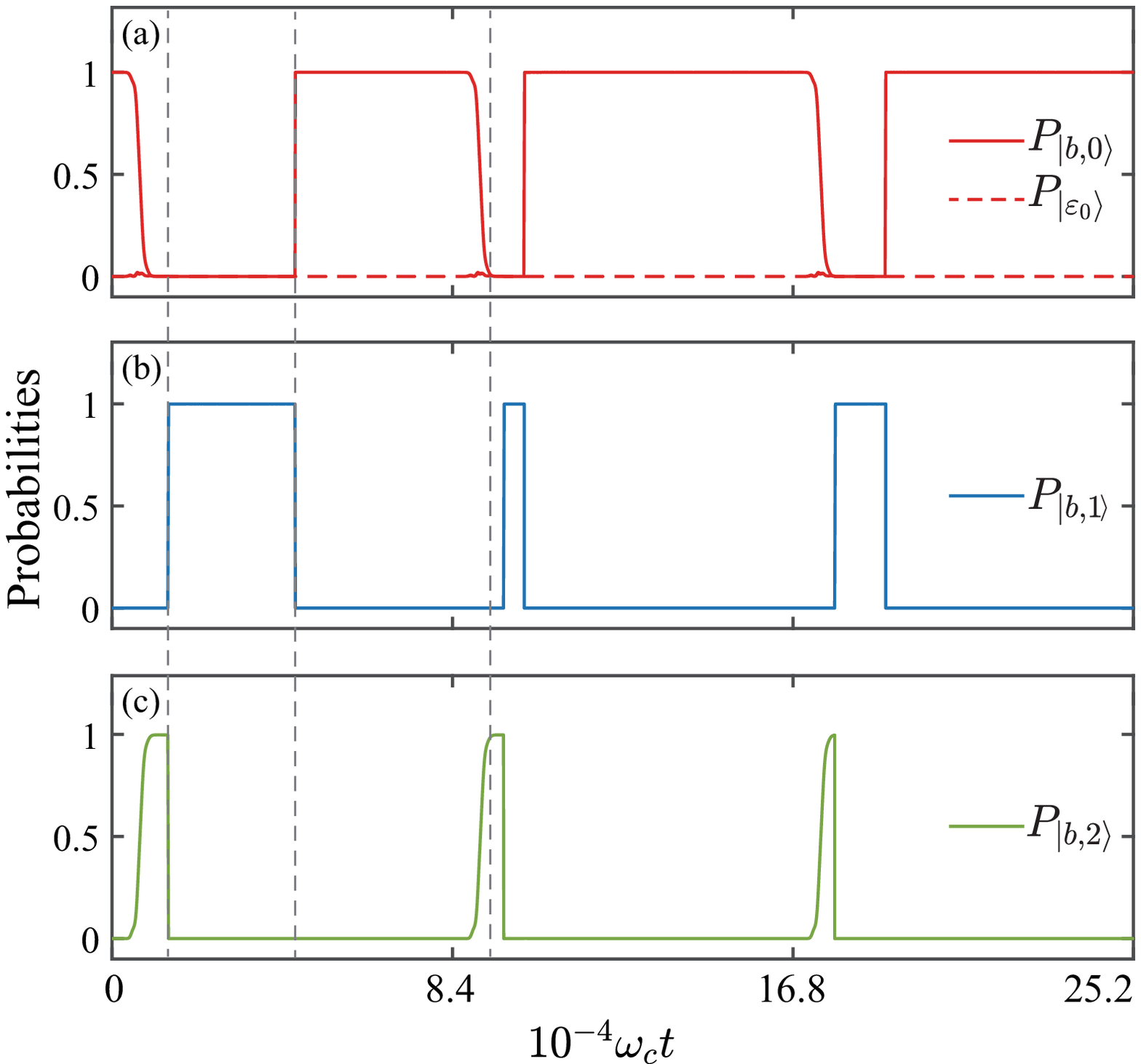}
\caption{(Color online) Quantum trajectories of the probabilities of the states $\vert b,j\rangle$ ($j=0,1,2$) and $\vert \varepsilon_{0}\rangle$ in the two-photon bundle emission. The parameters are the same as those in Figure~		\protect\ref{Fig3_two_photons_diss}.}
\label{Fig8_2_photons_mk}
\end{figure}
\noindent  of the state $\left\vert\varepsilon _{0}\right\rangle $ \textcolor{black}{($\left\vert \varepsilon_{1}\right\rangle$)} as functions of the time $t$ by solving eq.~(\ref{master}). 
Furthermore, to study the statistical characteristics of the emitted photons, we 
numerically calculate the generalized second-order photon correlation functions of the $
N$-photon bundle~\cite{Munoz2014,Bin2021}
\begin{equation}
g_{N}^{\left( 2\right) }\left( t,t+\tau \right) =\frac{\left\langle X^{\dag
		N}\left( t\right) X^{\dag N}\left( t+\tau \right) X^{N}\left( t+\tau \right)
	X^{N}\left( t\right) \right\rangle }{\left\langle X^{\dag N}\left( t\right)
	X^{N}\left( t\right) \right\rangle \left\langle X^{\dag N}\left( t+\tau
	\right) X^{N}\left( t+\tau \right) \right\rangle },  \label{gN_2tau}
\end{equation}%
where the operator $X$ is defined as 
\begin{equation}
X=\sum^{\infty}_{n,m>n}\left\langle \psi _{n}\right\vert ( a^{\dag }+a) \left\vert
\psi _{m}\right\rangle \left\vert \psi _{n}\right\rangle \left\langle \psi
_{m}\right\vert .
\end{equation}%
Note that for $N=1$, eq.~(\ref{gN_2tau}) gives the standard second-order correlation function, and at $\tau =0$,  eq.~(\ref{gN_2tau}) is the equal-time second-order correlation \textcolor{black}{function}
of $N$-photon bundle.
In the following, \textcolor{black}{we will focus on even- (two, four and six-photon) and odd-photon (three-photon) bundle emission separately to illustrate our approach.}

\subsection{Two-photon bundle}
We first consider the emission of two-photon bundle from the dissipative cavity.
In Figure~\ref{Fig3_two_photons_diss}(a), the Gaussian pulses of the two driving fields are presented over three cycles of the STIRAP process. The cycles are separated by the duration $T_1$. In Figure~\ref{Fig3_two_photons_diss}(b-d), we plot the probabilities $P_{\left\vert b,0\right\rangle
},\,P_{\left\vert b,1\right\rangle },\,P_{\left\vert b,2\right\rangle }$, and $%
P_{\left\vert \varepsilon _{0}\right\rangle }$ as functions of the normalized time $10^{-4} \omega_{c}t$. 
After applying the Gaussian pulse $\Omega_{2}\left( t\right) $ followed by the Gaussian pulse $\Omega _{1}\left( t\right)$, as shown in Figure~\ref{Fig3_two_photons_diss}(a), we find that the initial state $\left\vert b,0\right\rangle $ is effectively transferred to  $\left\vert b,2\right\rangle $ with a probability
\begin{figure}[tbp]
\center
\includegraphics[width=8.3cm]{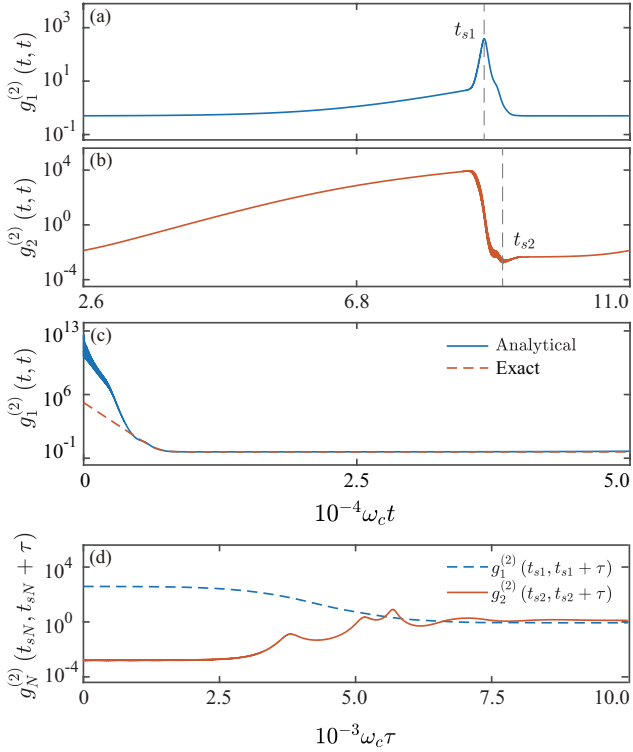}
\caption{(Color online) The equal-time and time-delayed second-order correlation functions for two-photon emission. (a) $g_{1}^{(2)}\left(t,t\right) $ and (b) $g_{2}^{(2)}\left( t,t\right) $ vs time $t$. The $t_{s1}$ ($t_{s2}$) is the time corresponding to the maximum (minimum) value of $g_{1}^{(2)}\left(t,t\right) $ [$g_{2}^{(2)}\left( t,t\right) $]. \textcolor{black}{(c) Analytical and \textcolor{black}{exact} solutions of the standard equal-time second-order correlation functions as a function of the scaled time $10^{-4}\omega_{c}t$ .} (d) $g_{N}^{(2)}\left(t_{sN},t_{sN}+\protect\tau \right) $ for $N$=1 and $N$=2 vs. the time delay $\tau$. The parameters are the same as those in Figure~\ref{Fig3_two_photons_diss}.}
\label{Fig5_two_photons_g2}
\end{figure}
\noindent   0.713 under the parameters $\lambda /\omega _{c}\!=\!0.6$, $\omega
_{b}/\omega _{c}\!=\!-6$, $\Omega _{1}/\omega _{c}\!=\!0.008$, $\Omega
_{2}/\Omega _{1}\!=\!6.8538$, \textcolor{black}{$\omega_{c}t_{1}=7960$, $\omega_{c}T=2200$} and 
$\kappa _{u}/\omega _{c}\!=\!0.0001$ ($u=a,\,ge,\,bg$). Because of finite cavity dissipation during the STIRAP, the state conversion probability is smaller than 1. 
The generated photons are then emitted to the cavity output by the decay processes $\left\vert b,2\right\rangle \rightarrow\left\vert b,1\right\rangle \rightarrow\left\vert b,0\right\rangle $, and 
the system returns to the initial state $\left\vert b,0\right\rangle $ after the emissions. The emission cycle repeats itself after a duration $T_{1}$, which needs to be sufficiently long to ensure the system returns to the initial state $\left\vert b,0\right\rangle$ before the start of the next emission cycle. 
Here the Gaussian pulses $\Omega _{l}\left( t\right) $ ($l=1,2$) satisfy the condition~(\ref{evo_con}) in order to achieve effective generation of the photon bundle. 

To study the dynamical emission of the photons, we simulate an initial quantum system by using the quantum jump approach~\cite{Plenio1998,Daley2014}.  In Figure~\ref{Fig8_2_photons_mk}(a-c), we plot the quantum trajectory of the probabilities of the states $\vert b,j\rangle$ ($j$=0,1,2) and $\vert \varepsilon_{0}\rangle$ starting from the initial state $\vert b,0\rangle$. 
Our result shows that after the STIRAP, the population in the state $\vert b,2\rangle$ is almost one as can be seen in  Figure~\ref{Fig8_2_photons_mk}(c). After the first photon is emitted out of the cavity, the system state collapses to $\vert b,1\rangle$ with a probability almost equal to 1, as shown in Figure~\ref{Fig8_2_photons_mk}(b). 
\begin{figure}[tbp]
\center
\includegraphics[width=8.3cm]{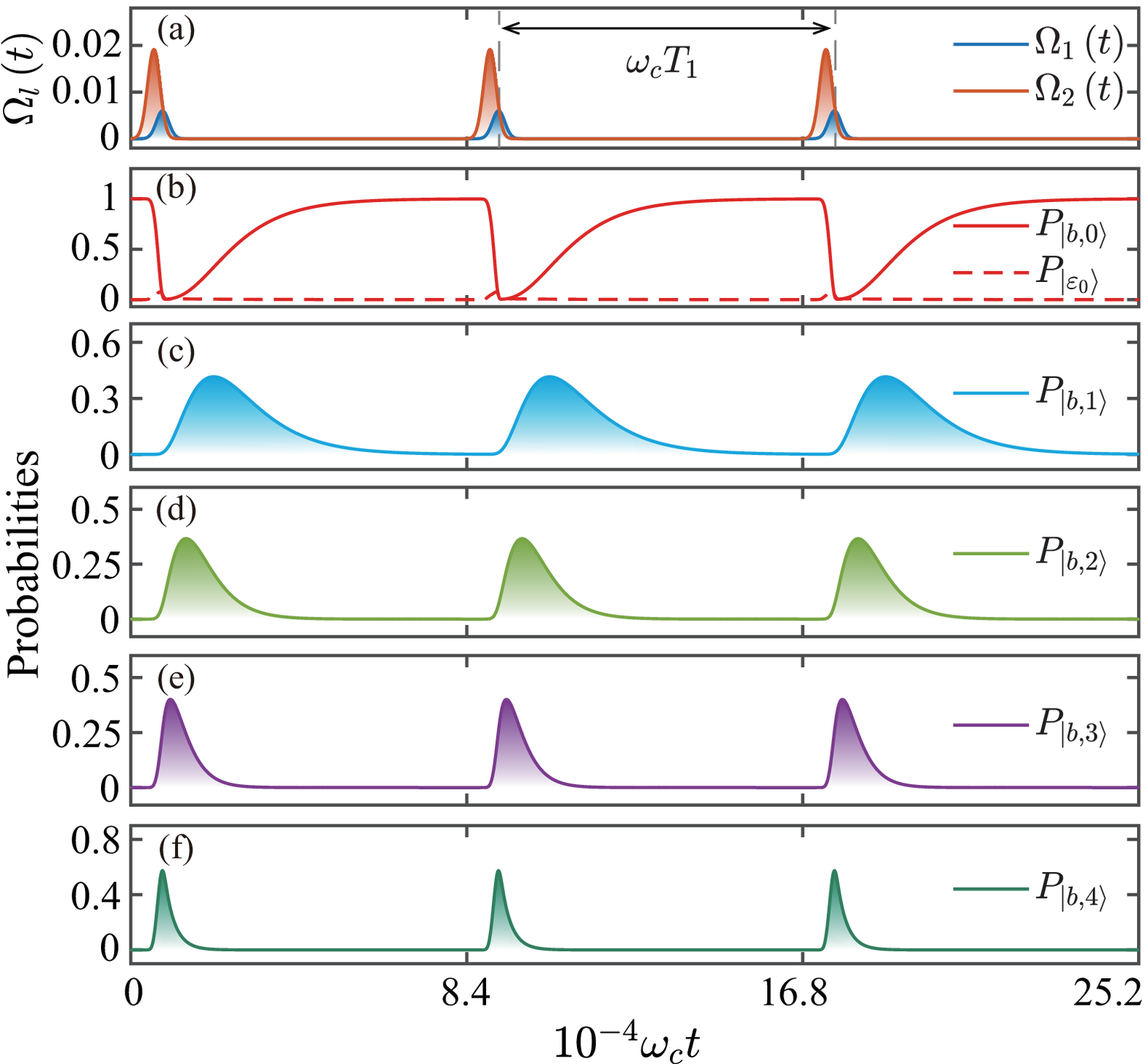}
\caption{(Color online) (a) The amplitudes $\Omega_{l}\left(t\right)$ ($l$
	=1,2) of the Gaussian pulses as a function of the scaled time $10^{-4}
	\protect\omega_{c}t$. (b)-(f) The probabilities $P_{\left\vert
		b,j\right\rangle }\left( t\right) $ ($j=0,1,\dotsb,4$) and $P_{\vert\protect
		\varepsilon_{0}\rangle}\left( t\right) $ vs. the scaled time. The decay rates are $\protect\kappa _{a}/\protect\omega _{c}=\protect\kappa _{ge}/\protect\omega _{c}=\protect\kappa 
	_{bg}/\protect\omega _{c}=0.0001$. Other parameters are the same as those in Figure~%
	\protect\ref{Fig2_two_four_photons_v1}(g)-(i).}
\label{Fig4_four_photons_diss_34}
\end{figure}
\noindent After the second photon is emitted, the system returns to the initial state $\vert b,0\rangle$ as shown in Figure~\ref{Fig8_2_photons_mk}(a). This result hence illustrates the two-photon bundle emission where the two photons are separated by a short temporal window determined by the cavity decay rate. 

To  investigate the statistical properties of the emitted photons, we numerically calculate the standard and generalized equal-time second-order correlation functions for $N$=1 and $N$=2 given by eq.~(\ref{gN_2tau}) at $\tau=0$. In Figure~\ref{Fig5_two_photons_g2}(a), we  plot the standard equal-time
second-order correlation function $g_{1}^{(2)}\left( t,t\right) $ within one emission 
cycle. We find that the maximum value of $g_{1}^{(2)}\left( t,t\right) $ at the time $t_{s1}$ is larger than one. This result implies that the photons are in a super-Poisson distribution with more than one photons emitted in the system. In Figure~\ref{Fig5_two_photons_g2}(b), we plot the generalized equal-time second-order correlation function $g_{2}^{(2)}\left( t,t\right) $ within one emission cycle. The minimum value of $g_{2}^{(2)}\left( t,t\right) $ at the time $t_{s2} $ is smaller than one, which corresponds to a sub-Poisson distribution of the emitted photon bundles. 

\textcolor{black}{\textcolor{black}{To analytically estimate} the correlation of \textcolor{black}{the} emitted photons, we assume that the system has been prepared \textcolor{black}{in} state $\vert b,2\rangle$ with the probability $\sin^{2}[\theta_{2}(t)]$ \textcolor{black}{during} the process of STIRAP, which is much slower than the cavity dissipation. The two cavity \textcolor{black}{photons} are emitted to the cavity output only through the dissipation of the cavity \textcolor{black}{mode} due to the decoupling of the atom from the cavity mode. In this case, the angle $\theta_{2}(t)$ can be treated as a constant during the cavity dissipation. Thus the  analytical solution of the equal-time  second-order
\begin{figure}[tbp]
	\center
	\includegraphics[width=8.3cm]{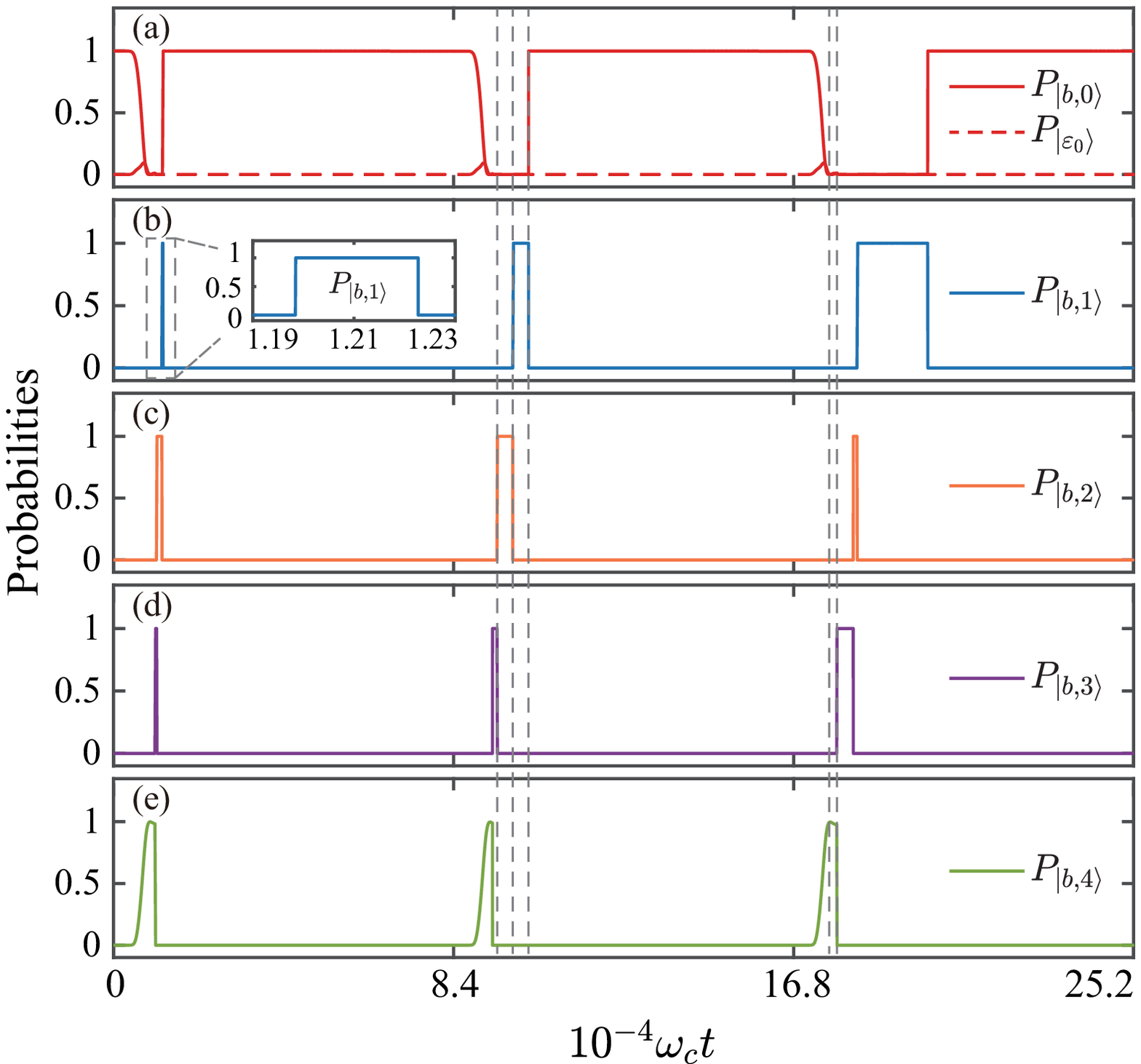}
	\caption{(Color online) Quantum trajectories of the probabilities of the states $\vert b,j\rangle$ $(j=0,1,\dotsb,4)$ and $\vert \varepsilon_{0}\rangle$ in the four-photon bundle emission. The parameters are the same as those in Figure~\ref{Fig4_four_photons_diss_34}.}
	\label{Fig8_4_photons_mk}
\end{figure}
\noindent  correlation function~\cite{Huang2014} can be \textcolor{black}{estimated} as
\begin{equation}
	g^{(2)}_{1}(t,t)=\frac{1}{2\sin[\theta_{2}(t)]}.\\
	~\label{g2tt_analysis}
\end{equation} We find that the correlation function~(\ref{g2tt_analysis}) is independent of \textcolor{black}{cavity dissipation}. Note that the analytical solution given in~(\ref{g2tt_analysis}) and the \textcolor{black}{exact} solution of the equal-time second order correlation function match well when $t\geqslant 4000$, as shown in Figure~\ref{Fig5_two_photons_g2}(c). \textcolor{black}{\textcolor{black}{Here} $t=4000$ corresponds to the beginning of the generation of two-photon state.} It is worth noting that the value of $g_{1}^{(2)}(t,t)$ is equal to 1/2 at $\theta_{2}(t)=\pi/2$ at the end of the STIRAP.}

\textcolor{black}{To further  investigate the statistical properties of the emitted photons, }we also calculate the time-delayed second-order correlation functions $g^{(2)}_{1}(t_{s1},t_{s1}+\tau)$ and $g^{(2)}_{2}(t_{s2},t_{s2}+\tau)$ as defined in eq.~(\ref{gN_2tau}).
Our result is given in Figure~\ref{Fig5_two_photons_g2}(d). 
It can be seen that $g_{1}^{\left( 2\right) }\left( t_{s1},t_{s1} \right)\!>\!g_{1}^{(2)}\left( t_{s1},t_{s1}+\tau \right) $ and $g_{2}^{\left( 2\right) }\left( t_{s2},t_{s2}\right) \!<\!g_{2}^{\left( 2\right) }\left(t_{s2},t_{s2}+\tau \right) $.
This result further confirms that the emitted photons are bunched while the 
two-photon bundles are antibunched. Hence our scheme can lead to the construction of a two-photon antibunched emitter. 

\textcolor{black}{\subsection{Four-photon and six-photon bundle}}
Next we study the generation of four-photon bundle by using the STIRAP technique. In Figure~\ref{Fig4_four_photons_diss_34}(a), the external Gaussian pulses $\Omega _{1}\left( t\right)$  and $\Omega _{2}\left( t\right)$ are given,  which satisfy the condition (\ref{evo_con}).  Similar to the studies for two-photon bundle,  we 
\begin{figure}[tbp]
\center
\includegraphics[width=8.3cm]{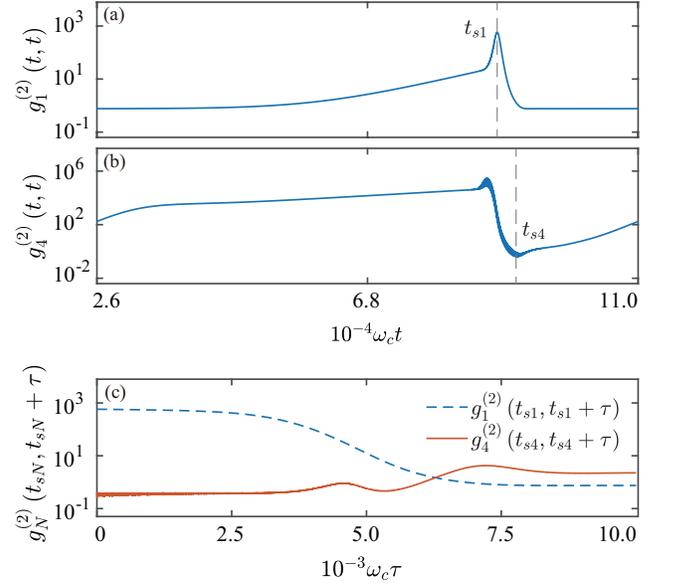}
\caption{(Color online) The equal-time and time-delayed second-order correlation functions for four-photon emission. (a) $g_{1}^{(2)}\left(t,t\right) $ and (b) $g_{4}^{(2)}\left( t,t\right) $ vs. time $t$. The $t_{s1}$ ($t_{s4}$) is the time correspoding to the maximum (minimum) value of $g_{1}^{(2)}\left(t,t\right) $ [$g_{4}^{(2)}\left( t,t\right) $]. (c) $g_{N}^{(2)}\left(t_{sN},t_{sN}+\protect\tau \right)$ for $N$=1 and $N$=4 vs. the time delay $\tau$. The parameters are the same as those in Figure~\ref{Fig4_four_photons_diss_34}.}
\label{Fig6_four_photons_g2_34}
\end{figure}
\noindent  calculate the probabilities $P_{\left\vert b,j\right\rangle }\left( t\right) $ on the state $%
\left\vert b,j\right\rangle $ $\left( j=0,1,\dotsb,4\right) $ and $%
P_{\left\vert \varepsilon _{0}\right\rangle }\left( t\right) $ on the 
state $\left\vert \varepsilon _{0}\right\rangle $, which are plotted in Figure~\ref{Fig4_four_photons_diss_34}(b-f). 
It can be shown that after the applied Gaussian pulses, the initial state $\left\vert
b,0\right\rangle $ is effectively transferred to the four-photon state $\left\vert b,4\right\rangle $ with a probability 0.575. The parameters used here are $\lambda/\omega _{c}\!=\!1.2,$ $\omega _{b}/\omega _{c}\!=\!-10,\Omega _{1}/\omega _{c}\!=\!0.006,$ $\Omega _{2}/\Omega _{1}\!=\!3.1814$, \textcolor{black}{$\omega_{c}t_{1}\!=\!7960$, $\omega_{c}T\!=\!2200$} and $\kappa
_{u}/\omega _{c}\!=\!0.0001$ ($u=a,\,ge,\,bg$). The cavity dissipation during the external pulses reduces the efficiency of the population transfer to be less than one. The generated photons will then decay to the cavity output one by one with $\left\vert b,4\right\rangle \rightarrow \left\vert
b,3\right\rangle \rightarrow \left\vert b,2\right\rangle \rightarrow
\left\vert b,1\right\rangle \rightarrow \left\vert b,0\right\rangle$, and the system will return to the initial state $\left\vert b,0\right\rangle $ after all photons are emitted. The cycle will repeat when the next set of Gaussian pulses are applied. We want to point out that the time interval $T_{1}$ of the cycles satisfies the condition:  $\kappa_{a}T_{1}\!\gg\!1 $, so that our photons will be released outside the cavity when the next cycle begins.

We use the quantum jump approach to obtain the quantum trajectory in this four-photon generation process. 
One quantum trajectory of the probabilities $P_{\vert b,j\rangle}(t)$ ($j=0,1,\dotsb,4$) and $P_{\vert \varepsilon_{0}\rangle}(t)$ is presented in Figure~\ref{Fig8_4_photons_mk}(a-e). It can be seen that four photons appear in the cavity after the applied Gaussian pulses, then the photons are emitted to the cavity output one by one with a short temporal window between adjacent output photons with the final state returning to the initial state $\vert b,0\rangle$. The result shows that the four photons are emitted in a 
\begin{figure}[tbp]
\center
\includegraphics[width=7.6cm]{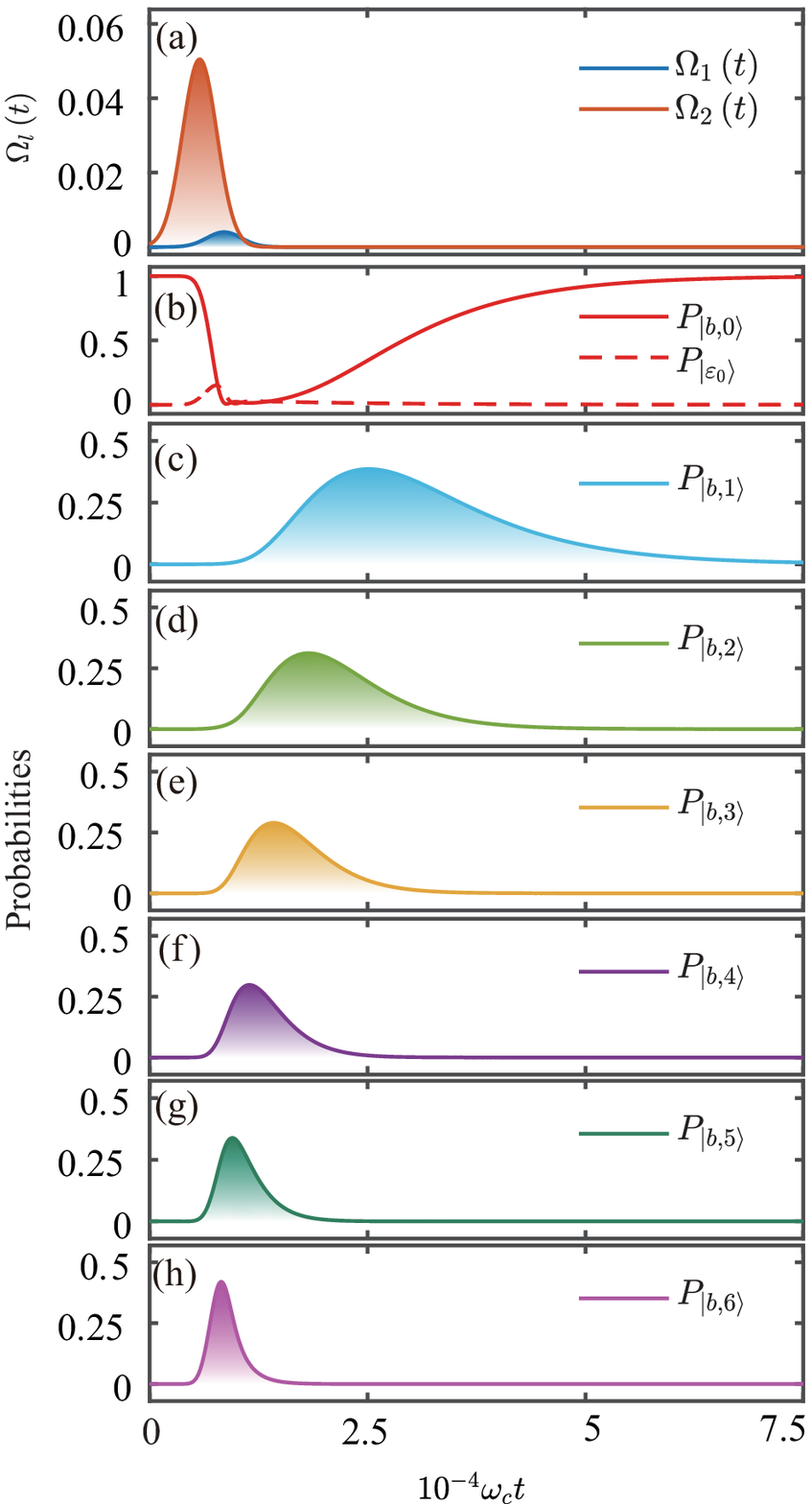}
\caption{(Color online)  \textcolor{black}{(a)} \textcolor{black}{The amplitudes $\Omega_{l}\left(t\right)$ ($l$
		=1,2) of the Gaussian pulses in one period  as a function of the scaled time $10^{-4}
		\protect\omega_{c}t$. (\textcolor{black}{b})-(h) The probabilities $P_{\left\vert
			b,j\right\rangle }\left( t\right) $ ($j=0,1,\dotsb,6)$ and $P_{\vert\protect
			\varepsilon_{0}\rangle}\left( t\right) $ vs. the scaled time.
		The decay rates in both cases are $\protect\kappa _{a}/\protect\omega _{c}=\protect\kappa _{ge}/\protect\omega _{c}=\protect\kappa_{bg}/\protect\omega _{c}=0.0001$. Other parameters are the same as those in Figure~\ref{Fig2_two_four_photons_v1}(j)-(l).}}
~\label{Fig10_three_six_photons}
\end{figure}
\noindent bundle before the next cycle begins.

To understand the statistical properties of the generated  photons, we calculate the equal-time second-order correlation functions for single photon $g^{\left(2\right)}_{1}\left(t,t\right)$ and four
photons $g^{\left(2\right)}_{4}\left(t,t\right)$, respectively, as given in Figure~\ref{Fig6_four_photons_g2_34}(a) and (b). 
We find that the maximum value of $g_{1}^{(2)}\left( t,t\right) $ at the time $t_{s1}$ is larger  than one, which shows that the generated photons are in a super-Poisson distribution at $t_{s1}$. Meanwhile, the minimum value of $g_{4}^{(2)}\left( t,t\right) $ at the time $t_{s4}$ is smaller than one,  which indicates that the four-photon bundles are in a sub-Poisson distribution at $t_{s4}$. To characterize the statistics of the emitted four-photon bundle, we further study the time-delayed second-order correlation functions $g^{\left(2\right)}_{1}\left(t_{s1},t_{s1}+\tau\right)$
and $g^{\left(2\right)}_{4}\left(t_{s4},t_{s4}+\tau\right)$ following the definition in eq.~(\ref{gN_2tau}).
The 
\begin{figure}[tbp]
\center
\includegraphics[width=7.6cm]{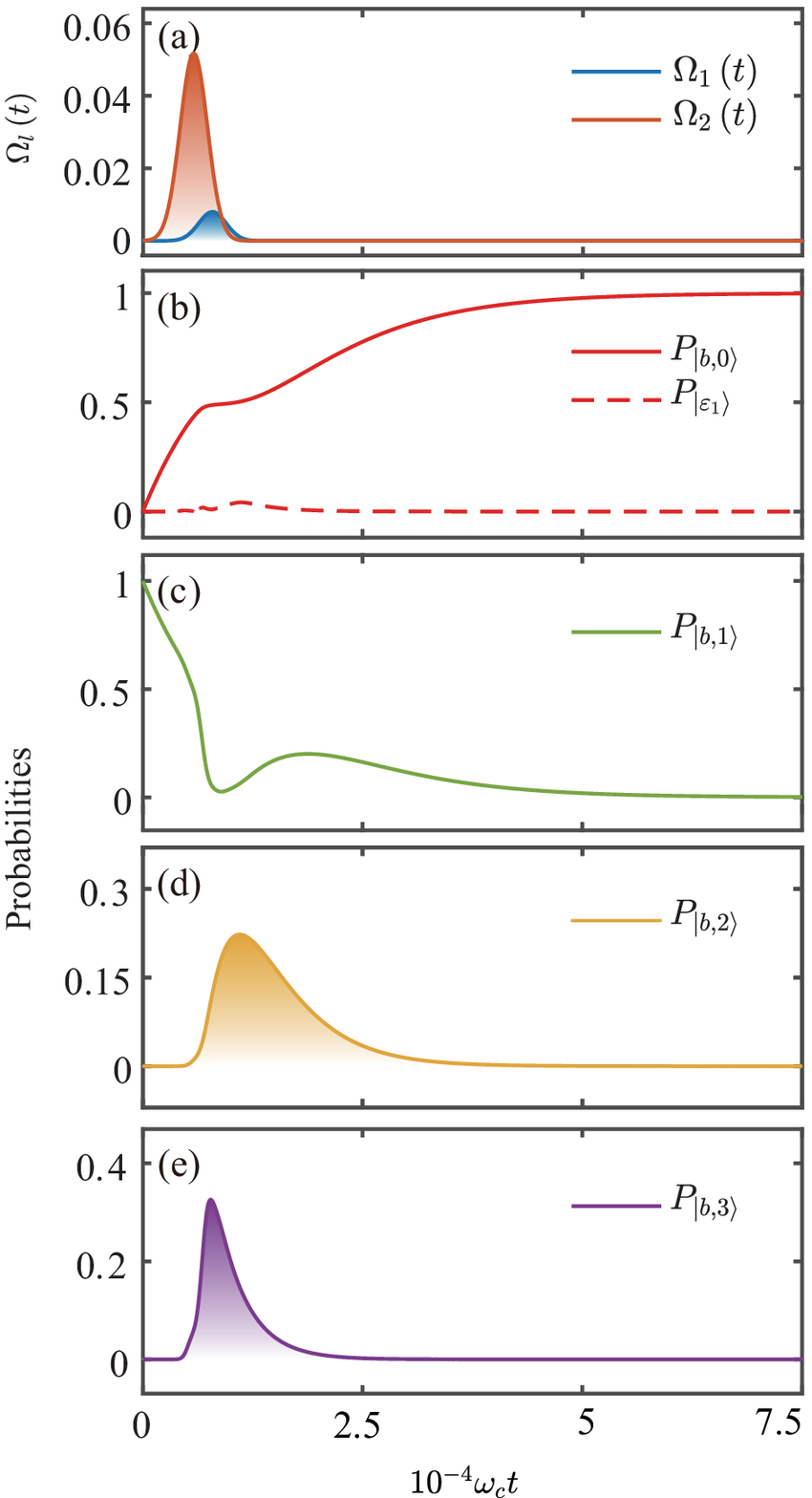}
\caption{(Color online)  \textcolor{black}{(a)} \textcolor{black}{The amplitudes $\Omega_{l}\left(t\right)$ ($l$
		=1,2) of the Gaussian pulses in one period  as a function of the scaled time $10^{-4}
		\protect\omega_{c}t$. (\textcolor{black}{b})-(e) The probabilities $P_{\left\vert
			b,j\right\rangle }\left( t\right) $ ($j=0,1,2,3$) and $P_{\vert\protect
			\varepsilon_{1}\rangle}\left( t\right) $ vs. the scaled time.  
		The decay rates in both cases are $\protect\kappa _{a}/\protect\omega _{c}=\protect\kappa _{ge}/\protect\omega _{c}=\protect\kappa_{bg}/\protect\omega _{c}=0.0001$. Other parameters are the same as those in Figure~\ref{Fig2_two_four_photons_v1}(d)-(f).}}
~\label{Fig11_three_photons}
\end{figure}
\noindent  numerical result of the time-delayed correlation functions is given in Figure~\ref{Fig6_four_photons_g2_34}(c). Similar to that of the two-photon bundle, $
g_{1}^{\left( 2\right) }\left( t_{s1},t_{s1}\right) >g_{1}^{(2)}\left(
t_{s1},t_{s1}+\tau \right)$, indicates bunched single-photon behavior, and 
$g_{4}^{\left( 2\right) }\left(
t_{s4},t_{s4}\right)< g_{4}^{\left( 2\right) }\left( t_{s4},t_{s4}+\tau
\right) $, indicates antibunched behavior for the four-photon bundles. 
This result verifies that the scheme gives a method to implement an antibunched four-photon bundle emitter.

\textcolor{black}{In order to demonstrate the emission of  multi-photon bundle clearly, we \textcolor{black}{also} study the generation of the six-photon bundle via the STIRAP technique. Under the driving pulses $\Omega_{1}\left(t\right)$ and $\Omega_{2}\left(t\right)$ given in Figure~\ref{Fig10_three_six_photons}(a), the one-period evolutions of  the populations $P_{\vert b,j\rangle}(t)$ on the state $\vert b,j\rangle$ ($j=0,1,\dotsb,6$) and $P_{\vert \varepsilon_{0}\rangle}(t)$ on the state $\vert \varepsilon_{0}\rangle$ are shown in Figure~\ref{Fig10_three_six_photons}(b)-(h). We can find that under the parameters $\lambda /\omega _{c}\!=\!1.2$, $\omega
_{b}/\omega _{c}\!=\!-10$, $\Omega _{1}/\omega _{c}\!=\!0.004$, $\Omega
_{2}/\Omega _{1}\!=\!12.6179$, $\omega_{c}t_{1}=8560$, $\omega_{c}T=2800$ and  $\kappa _{u}/\omega _{c}\!=\!0.0001$ ($u=a,\,ge,\,bg$),  the initial state $\vert b,0\rangle$ is effectively transferred to the six-photon state $\vert b,6\rangle$ with probability 0.421. The efficiency of the population transfer is smaller than one due to the cavity dissipation during the STIRAP. The photons on the state $\vert b,6\rangle$ will decay to the cavity output one by one with $\left\vert b,6\right\rangle \rightarrow\left\vert b,5\right\rangle \rightarrow\left\vert b,4\right\rangle \rightarrow \left\vert
b,3\right\rangle \rightarrow \left\vert b,2\right\rangle \rightarrow
\left\vert b,1\right\rangle \rightarrow \left\vert b,0\right\rangle$, and the system returns to the initial state $\vert b,0\rangle$ after emissions.  After applying the same Gaussian pulses again,  \textcolor{black}{another set of six photons} will be generated. The result shows that the generation of six-photon bundle can}
\textcolor{black}{be implemented in our scheme.} 

\textcolor{black}{\subsection{Three-photon bundle}}
\textcolor{black}{Next we study the generation of three-photon bundle \textcolor{black}{as an example of odd photon bundle}. \textcolor{black}{Its} physical process is similar to the case of even-photon bundles. We consider the excited state $\vert \varepsilon_{1}\rangle$ of the QRM in the case of three-photon bundle, and the system also is reduced to the three-level system by adjusting the frequencies of the driving pulses.  Under the driving pulses $\Omega _{1}\left( t\right)$  and $\Omega _{2}\left( t\right)$  given in Figure~\ref{Fig11_three_photons}(a), we plot the probabilities $P_{\vert b,j\rangle}$  ($j=0,1,2,3$) and $P_{\vert \varepsilon_{1}\rangle}$ within one period as functions of the normalized time $10^{-4}\omega_{c}t$, as shown in Figure~\ref{Fig11_three_photons}(b)-(e). Here, the initial state is chosen to the state $\vert b,1\rangle$ with the cavity field in one-photon state $\vert 1\rangle$ and the atom in the lowest state $\vert b\rangle$. The population transfers from initial state $\vert b,1\rangle$ to the three-photon state $\vert b,3\rangle$ with the probability 0.326 after applying Gaussian pulses. The parameters used here are $\lambda/\omega _{c}\!=\!0.6,$ $\omega _{b}/\omega _{c}\!=\!-6,\Omega _{1}/\omega _{c}\!=\!0.008,$ $\Omega _{2}/\Omega _{1}\!=\!6.4641$, $\omega_{c}t_{1}\!=\!7960$, $\omega_{c}T\!=\!2200$ and $\kappa_{u}/\omega _{c}\!=\!0.0001$ ($u=a,\,ge,\,bg$). 	It is worth noting that a little population on the state $\vert b,1\rangle$ is initially dissipated into the state $\vert b,0\rangle$ through the process of dissipation $\left\vert b,1\right\rangle \rightarrow\left\vert b,0\right\rangle$, so that the efficiency of the population transfer is smaller than one. With dissipation involved, the generated photons will be emitted to the cavity output one by one with $\left\vert b,3\right\rangle \rightarrow \left\vert b,2\right\rangle \rightarrow
\left\vert b,1\right\rangle \rightarrow \left\vert b,0\right\rangle$, and the system will be in the initial state after all dissipations. Due to the disappearance of the component $\vert g,0\rangle$ in the first excited state, we \textcolor{black}{prepare again the system in state $\vert b,1\rangle$, then repeat the emission cycle.} Thus, by choosing an appropriate initial state, the generation of three-photon \textcolor{black}{bundles} can be implemented in our scheme.}

\section{Conclusions and discussion}~\label{Dis_con}
We proposed a deterministic approach to generate multi-photon bundles via 
STIRAP for a $\Xi $-type atom coupled to a cavity mode in the ultrastrong or deep-strong coupling regime. By applying two appropriately designed Gaussian pulses, the system state can be transferred on-demand to a multi-photon state via the STIRAP technique. The photons will then decay to the cavity output in a bundle. By studying the quantum trajectory of the system and the standard and generalized second-order correlation functions, we find that the emitted single photons are in a super-Poisson distribution and the emitted multi-photon bundles are antibunched in a sub-Poisson distribution. This scheme provides a venue to implement efficient, on-demand multi-photon emitters.  

In this work, we discussed the generation of \textcolor{black}{multiple} photons based on the selection of the driving parameters in the calculation. We \textcolor{black}{showed that both even and} odd numbers of photon \textcolor{black}{bundles} can \textcolor{black}{be emitted on-demand by} adjusting the frequencies of the driving pulses to aim at \textcolor{black}{the ground state or} the first excited state of the QRM. 
Note that the successful generation of desired multi-photon bundles requires that the \textcolor{black}{duration} of the pulse cycle $T_1$ is sufficiently long with $\kappa_{a}T_{1}\gg 1$, so that the system can return to the proper initial state \textcolor{black}{$\left\vert b,M\right\rangle $ ($M$=0 or 1)} before the next cycle starts. 

The realization of our scheme is within reach of current state-of-the-art experimental technology. The
ultrastrong and deep-strong coupling regimes have been realized in several platforms, such as 
superconducting circuits \cite{Niemczyk2010,Forn2010,Yoshihara2017,Yoshihara2018}, intersubband polaritons~
\cite{Askenazi2017}, Landau polaritons~\cite{Bayer2017,Muravev2011}, organic molecules~\cite{Schwartz2011,Barachati2018}, and optomechanics~\cite{Benz2016}. Hence our method is
a practical approach that can lead to the construction of on-demand multi-photon sources.

\begin{acknowledgments}
Jin-Feng Huang was supported in part by the National Natural
Science Foundation of China (Grant No. 12075083), and Natural Science Foundation of Hunan Province, China (Grant No. 2020JJ5345). Lin Tian was supported by the National Science Foundation (Grant Nos. 2006076, and 2037987), and UC-MRPI Program (Grant No. ID M23PL5936).
\end{acknowledgments}



\begin{thebibliography}{99}
\bibitem{Niemczyk2010} T. Niemczyk, F. Deppe, H. Huebl, E. P. Menzel, F. Hocke, M. J. Schwarz, J. J. Garcia-Ripoll, D. Zueco, T. H{\"u}mmer, E. Solano, A. Marx, and R. Gross, Nat. Phys. \textbf{6}, 772 (2010).

\bibitem{Forn2010} P. Forn-D\'{ı}az, J. Lisenfeld, D. Marcos, J. J. Garc\'{ı}a-Ripoll, E. Solano, C. J. P. M. Harmans, and J. E. Mooij, Phys. Rev. Lett. \textbf{105}, 237001 (2010).

\bibitem{Schwartz2011} T. Schwartz, J. A. Hutchison, C. Genet, and T. W. Ebbesen, Phys. Rev. Lett. \textbf{106}, 196405 (2011).

\bibitem{Benz2016} F. Benz, M. K. Schmidt, A. Dreismann, R. Chikkaraddy, Y. Zhang, A. Demetriadou, C. Carnegie, H. Ohadi, B. de Nijs, R. Esteban, J. Aizpurua, and J. J. Baumberg, Science \textbf{354}, 726 (2016).

\bibitem{Askenazi2017} B. Askenazi, A. Vasanelli, Y. Todorov, E. Sakat, J.-J. Greffet, G. Beaudoin, I. Sagnes, and C. Sirtori, ACS photonics \textbf{4}, 2550 (2017).

\bibitem{Barachati2018} F. Barachati, J. Simon, Y. A. Getmanenko, S. Barlow, S. R.
Marder, and S. K\'{e}na-Cohen, ACS Photonics \textbf{5}, 119 (2018).

\bibitem{Yoshihara2017} F. Yoshihara, T. Fuse, S. Ashhab, K. Kakuyanagi, S. Saito, and K. Semba, Nat. Phys. \textbf{13}, 44 (2017).

\bibitem{Bayer2017} A. Bayer, M. Pozimski, S. Schambeck, D. Schuh, R. Huber,
D. Bougeard, and C. Lange, Nano Lett. \textbf{17}, 6340 (2017).

\bibitem{Yoshihara2018} F. Yoshihara, T. Fuse, Z. Ao, S. Ashhab, K. Kakuyanagi, S. Saito, T. Aoki, K. Koshino, and K. Semba, Phys. Rev. Lett. \textbf{120}, 183601 (2018).

\bibitem{Muravev2011} V. M. Muravev, I. V. Andreev, I. V. Kukushkin, S. Schmult, and W. Dietsche, Phys.
Rev. B \textbf{83}, 075309 (2011).

\bibitem{Nataf2010} P. Nataf and C. Ciuti, Phys. Rev. Lett. \textbf{104}, 023601 (2010).

\bibitem{Ridolfo2012} A. Ridolfo, M. Leib, S. Savasta, and M. J. Hartmann, Phys. Rev. Lett. \textbf{109}, 193602 (2012).

\bibitem{Wang2012} Z. H. Wang, Y. Li, D. L. Zhou, C. P. Sun, and P. Zhang, Phys. Rev. A \textbf{86}, 023824 (2012).

\bibitem{Shi2018} T. Shi, Y. Chang, and J. J. Garc\'{ı}a-Ripoll, Phys. Rev. Lett. \textbf{120}, 153602 (2018).

\bibitem{Hwang2015} M.-J. Hwang, R. Puebla, and M. B. Plenio, Phys. Rev. Lett. \textbf{115}, 180404 (2015).

\bibitem{Chen2021} X. Chen, Z. Wu, M. Jiang, X.-Y. L\"{u}, X. Peng, and J. Du, Nat. Commun. \textbf{12}, 6281 (2021).

\bibitem{Cai2021} M.-L. Cai, Z.-D. Liu, W.-D. Zhao, Y.-K. Wu, Q.-X. Mei, Y. Jiang, L. He, X. Zhang, Z.-C. Zhou, and L.-M. Duan, Nat. Commun. \textbf{12}, 1126 (2021).

\bibitem{Chenye2022} Y.-H. Chen, A. Miranowicz, N. Lambert, W. Qin, R. Stassi, Y. Xia, S.-B. Zheng and F. Nori, arXiv: 2207.12156.

\bibitem{Garziano2015} L. Garziano, R. Stassi, V. Macr\`{ı}, A. F. Kockum, S. Savasta, and F. Nori, Phys. Rev. A \textbf{92}, 063830 (2015).

\bibitem{Ma2020} K. K. W. Ma, Phys. Rev. A \textbf{102}, 053709 (2020).

\bibitem{Huang2015} J.-F. Huang and C. K. Law, Phys. Rev. A \textbf{91}, 023806 (2015).

\bibitem{Huang2017} J.-F. Huang, J.-Q. Liao, L. Tian, and L.-M. Kuang, Phys. Rev. A \textbf{96}, 043849 (2017).

\bibitem{Huang2020} J.-F. Huang, J.-Q. Liao, and L.-M. Kuang, Phys. Rev. A \textbf{101}, 043835 (2020).

\bibitem{Huang2022} J.-F. Huang and L. Tian, arXiv:2208.12524. 

\bibitem{Stassi2013} R. Stassi, A. Ridolfo, O. Di Stefano, M. J. Hartmann, and S. Savasta, Phys. Rev. Lett. \textbf{110}, 243601 (2013).

\bibitem{Huang2014} J.-F. Huang and C. K. Law, Phys. Rev. A \textbf{89}, 033827 (2014).

\bibitem{Cirio2016} M. Cirio, S. De Liberato, N. Lambert, and F. Nori, Phys. Rev. Lett. \textbf{116}, 113601 (2016).

\bibitem{Munoz2014} C. S. Mu\~{n}oz, E. del Valle, A. G. Tudela, K. M \"{u}ller, S. Lichtmannecker, M. Kaniber, C. Tejedor, J. J. Finley, and F. P. Laussy, Nat. Photonics \textbf{8}, 550 (2014).

\bibitem{Walther2006} H. Walther, B. T. H. Varcoe, B.-G. Englert, and T. Becker, Rep. Prog. Phys. \textbf{69}, 1325
(2006).

\bibitem{Brien2009} J. L. O’Brien, A. Furusawa, and J. Vu\v{c}kovi\'{c}, Nat. Photonics \textbf{3}, 687 (2009).

\bibitem{Giovannetti2004} V. Giovannetti, S. Lloyd, and L. Maccone, Science \textbf{306}, 1330 (2004).

\bibitem{Giovannetti2006}  V. Giovannetti, S. Lloyd,  and  L. Maccone, Phys. Rev. Lett. \textbf{96}, 010401 (2006).

\bibitem{Angelo2001} M. D’Angelo, M. V. Chekhova, and Y. Shih, Phys. Rev. Lett. \textbf{87}, 013602 (2001).

\bibitem{Kimble2008} H. J. Kimble, Nature (London) \textbf{453}, 1023 (2008).

\bibitem{Ball2011} P. Ball, Nature (London) \textbf{474}, 272 (2011).

\bibitem{Sim2012} N. Sim, M. F. Cheng, D. Bessarab, C. M. Jones, and L. A. Krivitsky, Phys. Rev. Lett. \textbf{109}, 113601 (2012).

\bibitem{Denk1990} W. Denk, J. H. Strickler, and W. W. Webb, Science \textbf{248}, 73 (1990).

\bibitem{Horton2013} N. G. Horton, K. Wang, D. Kobat, C. G. Clark, F. W. Wise, C. B. Schaffer, and C. Xu,  Nat. Photonics \textbf{7}, 205 (2013).

\bibitem{Bienias2014} P. Bienias, S. Choi, O. Firstenberg, M. F. Maghrebi, M. Gullans, M. D. Lukin, A. V. Gorshkov, and H. P. B\"{u}chler, Phys. Rev. A \textbf{90}, 053804 (2014).

\bibitem{Maghrebi2015} M. F. Maghrebi, M. J. Gullans, P. Bienias, S. Choi, I. Martin, O. Firstenberg, M. D. Lukin, H. P. B\"{u}chler, and A. V. Gorshkov, Phys. Rev. Lett. \textbf{115}, 123601 (2015).

\bibitem{Liao2010} J.-Q. Liao and C. K. Law, Phys. Rev. A \textbf{82}, 053836 (2010).

\bibitem{Miranowicz2013} A. Miranowicz, M. Paprzycka, Y.-x. Liu, J. c. v. Bajer, and F. Nori, Phys. Rev. A \textbf{87}, 023809 (2013).

\bibitem{Dousse2010} A. Dousse, J. Suffczy\'{n}ski, A. Beveratos, O. Krebs, A. Lema\^{ı}tre, I. Sagnes, J. Bloch, P. Voisin, and P. Senellart,  Nature (London) \textbf{466}, 217 (2010).

\bibitem{Ota2011} Y. Ota, S. Iwamoto, N. Kumagai, and Y. Arakawa, Phys. Rev. Lett. \textbf{107}, 233602 (2011).

\bibitem{Muller2014} M. M\"{u}ller, S. Bounouar, K. D. J\"{o}ns, M. Gl\"{a}ssl, and P. Michler, Nat. Photonics \textbf{8}, 224 (2014).

\bibitem{Chang2016} Y. Chang, A. Gonz\'{a}lez-Tudela, C. S\'{a}nchez Mu\~{n}oz, C. Navarrete-Benlloch, and T. Shi, Phys. Rev. Lett. \textbf{117}, 203602 (2016).


\bibitem{Strekalov2014} D. V. Strekalov, Nat. Photonics \textbf{8}, 500 (2014).

\bibitem{SanchezMunoz2018} C. S. M. noz, F. P. Laussy, E. del Valle, C. Tejedor, and A. Gonz\'{a}lez-Tudela, Optica \textbf{5}, 14 (2018).

\bibitem{Bin2021} Q. Bin, Y. Wu, and X.-Y. L\"{u}, Phys. Rev. Lett. \textbf{127}, 073602 (2021).

\bibitem{Deng2021} Y. G. Deng, T. Shi, and S. Yi, Photonics Res. \textbf{9}, 1289 (2021).

\bibitem{Ma2021} S.-l. Ma, X.-k. Li, Y.-l. Ren, J.-k. Xie, and F.-l. Li, Phys. Rev. Research \textbf{3}, 043020 (2021).

\bibitem{Gonz2015} A. Gonz\'{a}lez-Tudela, V. Paulisch, D. E. Chang, H. J. Kimble, and J. I. Cirac, Phys. Rev. Lett. \textbf{115}, 163603 (2015).

\bibitem{Douglas2016} J. S. Douglas, T. Caneva, and D. E. Chang, Phys. Rev. X \textbf{6}, 031017 (2016).

\bibitem{Gonz2017} A. Gonz\'{a}lez-Tudela, V. Paulisch, H. J. Kimble, and J. I. Cirac, Phys. Rev. Lett. \textbf{118}, 213601 (2017).

\bibitem{Bergmann1998} K. Bergmann, H. Theuer, and B. W. Shore, Rev. Mod. Phys. \textbf{70}, 1003 (1998).

\bibitem{Vitanov2017} N. V. Vitanov, A. A. Rangelov, B. W. Shore, and K. Bergmann, Rev. Mod. Phys. \textbf{89}, 015006 (2017).


\bibitem{Liu2022} Y.-H. Liu, X.-L. Yin, J.-F. Huang, and J.-Q. Liao, Phys. Rev. A \textbf{105}, 023504 (2022).

\bibitem{Rabi1936} I. I. Rabi, Phys. Rev. \textbf{49}, 324 (1936).

\bibitem{Braak2011} D. Braak, Phys. Rev. Lett. \textbf{107}, 100401 (2011).

\bibitem{breuer2002} H.-P. Breuer and F. Petruccione, \emph{The theory of open quantum systems} (Oxford University Press, New York, 2002).

\bibitem{Plenio1998} M. B. Plenio and P. L. Knight,  Rev. Mod. Phys. \textbf{70},
101 (1998).

\bibitem{Daley2014} A. J. Daley, Adv. Phys. \textbf{63}, 77 (2014).

\end{thebibliography}
%


\end{document}